\documentclass[iop]{emulateapj}

\shorttitle{Metallicity distribution functions of the MCs}
\shortauthors{Haschke et al.}

\begin{document}
\submitted{Accepted for publication in Astronomical Journal}
   \title{Metallicity distribution functions of the old populations of the Magellanic Clouds from RR\,Lyrae stars}


\author{Raoul Haschke,
	  Eva K. Grebel,
          Sonia Duffau,
	  and
	  Shoko Jin\footnote{Alexander von Humboldt research fellow}}

\email{haschke@ari.uni-heidelberg.de}

\affil{Astronomisches Rechen-Institut, Zentrum f\"ur Astronomie der Universit\"at Heidelberg,
	      M\"onchhofstra\ss e 12-14, D-69120 Heidelberg, Germany}\

\begin{abstract}
We present the first metallicity distribution functions of the old field populations in the Magellanic Clouds. Our metallicities are based on the Fourier decomposition of Type\,ab RR\,Lyrae light curves from the Optical Gravitational Lensing Experiment (OGLE-III).  On the metallicity scale of Zinn \& West; we find a mean metallicity of $\mathrm{[Fe/H]} = -1.50 \pm 0.24$\,dex based on 16776 RR\,Lyrae stars in the Large Magellanic Cloud (LMC). For the Small Magellanic Cloud (SMC) we obtain $-1.70 \pm 0.27$\,dex based on 1831 RR\,Lyrae stars. These uncertainties represent the intrinsic spread in the population rather than the standard deviation of the mean.\\\noindent\hspace*{1em}
Our results are in good agreement with the few existing spectroscopic metallicity determinations for LMC RR\,Lyrae stars from the literature. For both the LMC and the SMC the metallicity spread exceeds 1\,dex in [Fe/H].  The distribution of metallicities in both Clouds is very uniform, and no significant metallicity gradient is detectable.  We also do not find any pronounced populations of extremely metal-poor RR\,Lyrae candidates with metallicities well below $-2$\,dex, although we need to caution that the photometric method used may overestimate the metallicities of metal-deficient stars. Moreover, because of stellar evolutionary effects one does not expect to observe many RR\,Lyrae stars among very metal-poor horizontal branch stars. We suggest that the Magellanic Clouds experienced fairly rapid and efficient early enrichment involving pre-enriched gas as well as possibly gas infall, while metal loss through outflows does not seem to have played a significant role. Moreover we suggest that the differences in the metallicities of the old population of LMC and SMC make an origin from a single, common progenitor unlikely, unless the separation happened very early on.
\end{abstract}


%

\section{Introduction}


\defcitealias{Jurcsik95}{J95}
\defcitealias{Zinn84}{ZW84}
\defcitealias{Dorfi99}{DF99}
\defcitealias{Deb10}{DS10}
\defcitealias{Feast10}{FAW10}

The origin and nature of the Magellanic Clouds (MCs) pose an interesting puzzle. They have long been known as the closest irregular satellites of the Milky Way, although galaxy surveys reveal that star-forming satellites in close proximity to massive galaxies are rare \citep[e.g.,][]{James11}, while very accurate {\em Hubble Space Telescope}-based determinations of the proper motions of the MCs \citep{Kallivayalil06a, Piatek08} have evoked a new discussion of their orbits \citep[e.g.,][]{Besla07, Besla10}. With recent new measurements of lower proper motions by \citet{Costa11} and \citet{Vieira10} the discussion has been further intensified. In particular, these studies have raised questions of whether the MCs might be on their first passage around the Milky Way (e.g., \citealt{Lux10} and references cited above; but see also, e.g., \citealt{Diaz11}). Similarly, it remains unclear whether the MCs were formed together, whether the Small Magellanic Cloud (SMC) might have been captured by the Large Magellanic Cloud (LMC) later on, or whether interactions or mergers might be responsible for the chemistry and peculiar structure of the SMC \citep[e.g.,][]{Mathewson88, Kallivayalil06b, Costa09, Tsujimoto09}.  \\\noindent\hspace*{1em}
One of the possible avenues of exploring the early history of the MCs relies on the study of their old populations. Generally, both globular clusters as well as field stars are suitable tracers of old populations.  The LMC does indeed contain a number of old globular clusters that are similarly old as the oldest globulars in other Galactic satellites and in the Milky Way \citep[e.g.,][]{Brocato96, Olsen98, Grebel04}.  In terms of their chemical composition, the LMC's globular clusters span a similar range of metallicities and show similar [$\alpha$/Fe] ratios as stars in the Galactic halo \citep[e.g.,][]{Johnson06, Mucciarelli10}.  In contrast, the SMC only contains one old globular cluster, NGC\,121, which is a few Gyr younger than the oldest Galactic and LMC globular clusters \citep{Glatt08a}. Thus one would ideally wish to turn to more numerous tracers of old populations such as those potentially present among field stars.  However, genuinely old field stars are difficult to identify in galaxies that experienced long-lasting star formation such as the MCs \citep[e.g.,][]{Holtzman99, Gallart08, Sabbi09}, since stellar populations of many different ages are then superimposed, rendering otherwise popular stellar tracers such as red giants rather ambiguous indicators \citep[see, e.g., Figure\,10 in][]{Koch06}. Using star clusters with spectroscopic metallicities and ages from main-sequence fitting, \citet{Glatt08b} showed that the SMC exhibits a considerable spread in metallicity at any given age. The study of evolutionary histories is further exacerbated if the galaxy in question has a non-negligible depth extent as appears to be the case for the SMC. Under such circumstances, using easily identifiable, special types of stars as tracers is a useful approach \citep[see, e.g.,][Figure\,3]{Grebel97}. One of these evolutionary tracers for evaluating the parameters of the old field population are RR\,Lyrae variables. With ages of at least 9\,Gyr \citep[e.g.][]{Sarajedini06}, they allow us to explore the early stages of the evolutionary history of a galaxy. \\\noindent\hspace*{1em}
When well-defined light curves are available, RR\,Lyrae stars can easily be identified. Apart from their distance and spatial distribution, their metallicity is a parameter of particularly high interest. Either spectroscopic measurements or Fourier decomposition of their light curves can be used to deduce the metallicity [Fe/H] \citep{Kovacs95, Jurcsik96}. \\\noindent\hspace*{1em}
Old populations including RR\,Lyrae stars tend to be widely distributed and often form the most extended stellar populations in galaxies.  This provides us with the possibility to not only measure the mean metallicity of genuinely old stars but also to obtain the metallicity spread, to derive detailed metallicity distribution functions (MDFs), and to identify possible metallicity gradients.  Overall, such studies can provide a fairly detailed picture of the chemical evolution that a galaxy experienced at early times.\\\noindent\hspace*{1em}
\citet{Butler82} first used the $\Delta$S method \citep{Preston59} to investigate the metallicity of six RR\,Lyrae stars in the LMC and ten RR\,Lyrae stars in the SMC. They found a significant difference between the two samples, with a mean metallicity of $\mathrm{[Fe/H]} = -1.4 \pm 0.1$\,dex for the LMC and $-1.8 \pm 0.2$\,dex for the SMC. In more recent studies, spectroscopic metallicities of old stars in the LMC were determined for sets of 74 to 98 RR\,Lyrae stars \citep[see Table\,\ref{table_FeH_literature}]{Gratton04, Borissova04, Borissova06}, while for the SMC no further spectroscopic data are available. These spectroscopic studies, which are concentrated close to the bar of the LMC, revealed mean metallicities of about $\mathrm{[Fe/H]} = -1.50$\,dex. \\\noindent\hspace*{1em}
With microlensing surveys such as the Optical Gravitational Lensing Experiment (OGLE) \citep{Udalski92, Udalski97, Udalski08a}, carefully chosen, extended regions with high stellar densities have been regularly monitored. In the Magellanic Clouds, OGLE focussed on the dense central regions. These surveys are well suited for finding variable stars, such as RR\,Lyrae stars. Thanks to the large number of repeat observations, variability properties of these stars can be determined with high precision. The parameters of the observed light curves of RR\,Lyrae stars of type\,ab can be used to determine the metallicity as shown first by \citet{Kovacs95}. \citet{Jurcsik96} presented a formalism to derive metallicities as a function of the period and the Fourier phase $\phi_{31}$. For cases where the Fourier parameters cannot be determined, for instance \citet{Brown04} and \citet{Sarajedini06} presented equations to infer the metallicity using only the period. These approaches have larger uncertainties than the equations presented by \citet{Jurcsik96} and are not used in our study.\\\noindent\hspace*{1em}
In recent years, the public data of the OGLE survey have been used to determine the MDF and the metallicity gradient of the MCs with RR\,Lyrae stars. The OGLE\,II data were used by \citet{Smolec05} to investigate the metallicity distribution of 5451 LMC RR\,Lyrae stars and the differences in [Fe/H] for Blazhko and non-Blazhko stars. For the non-Blazhko RR\,Lyrae stars a mean metallicity of $\mathrm{[Fe/H]} = -1.218 \pm 0.004$\,dex was determined, while the Blazhko stars had a mean value of $\mathrm{[Fe/H]} = -1.28 \pm 0.01$\,dex. \citet{Deb10} used the OGLE\,II data of the SMC to determine metallicities for 335 RR\,Lyrae stars of type\,ab and found a mean metallicity of $\mathrm{[Fe/H]} = -1.56 \pm 0.25$\,dex. In \citet[from now on FAW10]{Feast10} a gradient of $0.01 \pm 0.002$\,dex kpc$^{-1}$ was inferred for the RR\,Lyrae stars of the LMC, observed by OGLE\,III. \\\noindent\hspace*{1em}
We use the entire sample of RR\,Lyrae stars provided by the third data release of the OGLE collaboration to calculate individual metallicities for each star of the MCs. We discuss the origin of our data in Section\,\ref{data}. In Section\,\ref{FeH_method} we present the method for calculating metallicities used in this paper. The results for the LMC are given in Section\,\ref{FeH_LMC} and for the SMC in Section\,\ref{FeH_SMC}. We thus obtain an MDF of the old population and investigate the existence/absence of a metallicity gradient for both MCs. We test whether the photometric metallicities lead to similar results as spectroscopic values in Section\,\ref{FeH_spec_phot} and compare different methods to estimate the metallicity from photometric data in Section\,\ref{phi31toV}. Section\,\ref{Summary} discusses the results obtained.
%
%
\section{Data}
\label{data}

In 1992 the OGLE experiment started its first phase of observations of the MCs using the 1\,m Swope telescope of Las Campanas Observatory, Chile \citep{Udalski92}. This telescope was replaced by a 1.3\,m telescope owned by the OGLE collaboration at Las Campanas in 1995, with which the second phase of the OGLE experiment was performed. The CCD camera of the first two phases had a field of view of $15' \times 15'$ and $2048 \times 2048$\,pixels \citep{Udalski97}. Data were taken preferentially in the $I$ band and additionally in the $V$ and $B$ bands until 2001, when the third phase of OGLE was started. The eight CCDs of the OGLE\,III camera, each with $2048 \times 4096$ pixels, took data until 2009. With a total of $8192 \times 8192$\,pixels, this camera covered a total area of $35' \times 35'$ \citep{Udalski03}. In the LMC nearly 40 square degrees were monitored, while in the SMC 14 square degrees were observed using $V$ and $I$ band filters. Altogether photometric data of high precision are provided for 35\,million stars in the LMC and 6.2\,million stars in the SMC\footnote{The catalogues are available from \url{http://ogle.astrouw.edu.pl/}}. In addition to the full photometric catalogue, subsets of interesting objects such as RR\,Lyrae stars or Cepheids have also been compiled by the OGLE collaboration.\\\noindent\hspace*{1em}
In the LMC the OGLE collaboration identified 17693 RR\,Lyrae stars of type\,ab \citep{Soszynski08}, while in the SMC only 1933 RR\,Lyrae stars of type\,ab were found \citep{Soszynski10b}. As well as light curve parameters, mean magnitudes in the $V$ and $I$ bands and periods were determined by OGLE. The $I$ band light curves were decomposed using a fifth order Fourier series following \citet{Simon93} and the Fourier amplitudes and phases determined.
%
%
\section{Method of metallicity determinations}
\label{FeH_method}
Determining metallicities of a sample of this size is currently only feasible via photometric estimates. We therefore use the techniques of metallicity estimates based on Fourier-decomposed light curve parameters. \citet{Simon81a} introduced this method for variable stars, using the first few orders of a Fourier decomposition, 
\begin{equation}
V = A_0 + \sum_{k=1}^{N}A_k\sin(k\omega t + \phi_k)
\label{Fourier}
\end{equation}
where $V$ is the observed magnitude, $A_{0/k}$ is the amplitude, $\phi_{k}$ the phase, $\omega$ the angular frequency, and $t$ the specific time at which a data point was measured. $N$ is the order to which the Fourier series is developed.\\\noindent\hspace*{1em}
The Fourier parameters of the different orders of the decomposition contain information about important physical properties of the variables. While the amplitude $A_k$ is a measure of the skewness, the phase $\phi_k$ corresponds to the acuteness of the light curve \citep{Stellingwerf87}. \citet{Simon93} showed how the Fourier parameters $\phi_{31} = \phi_{3} - 3\phi_{1}$ are related to temperature and luminosity, while \citet{Kovacs95} presented the close connection between the period, the Fourier parameter $\phi_{31}$ and the metallicity of RR\,Lyrae stars. The relation for a sinusoidal Fourier decomposition of a $V$ band light curve is given by Equation\,(3) of \citet{Jurcsik96}. \\\noindent\hspace*{1em}
Fourier parameters differ depending on the filter in which the light curve is observed. As the OGLE photometry is mostly done in the $I$ band, the $V$ band equation by \citet{Jurcsik96} cannot be used. \citet{Smolec05} recalibrated the relation to be suitable for $I$ band photometry. Combining light curve parameters of 28 RR\,Lyrae stars with their spectroscopic metallicities ranging from $-1.71\,\mathrm{dex} < \mathrm{[Fe/H]}_{\mathrm{J95}} < +0.08$\,dex, on the metallicity scale of \citet[from now on J95]{Jurcsik95}, \citet{Smolec05} found the linear equation
\begin{eqnarray}
\mathrm{[Fe/H]}_{\mathrm{J95}} & = & (-3.142 \pm 0.646) - \label{FeH_equation}\\
& & (4.902 \pm 0.375)\,P + (0.824 \pm 0.104)\,\phi_{31}, \nonumber
\end{eqnarray}
with $\phi_{31}$ being the Fourier phase and $P$ being the period of the RR\,Lyrae star.  We adopt a slightly larger metallicity range of $ -2.0\,\mathrm{dex} < \mathrm{[Fe/H]} < +0.2$\,dex as a presumably reliable domain. This larger range is the metallicity range given by the calibration in \citet{Smolec05} plus the mean uncertainty of the method ($\pm 0.18$\,dex). \\\noindent\hspace*{1em} 
The \citetalias{Jurcsik95} metallicity scale, which is the basis for the relation by \citet{Smolec05}, leads to higher metallicities compared with other scales. \citet{Gratton04} find that the \citetalias{Jurcsik95} scale is more metal-rich by 0.28\,dex than the scale of \citet[later on ZW84]{Zinn84}, in agreement with \citet{Sandage04b} and \citet{Clementini05}. Using six globular clusters with metallicities between $-2.25\,\mathrm{dex} < \mathrm{[Fe/H]} < -1.0$\,dex, \citet{Papadakis00} derived the transformation relation
\begin{equation}
\mathrm{[Fe/H]}_{\mathrm{ZW84}} = 1.028\,\mathrm{[Fe/H]}_{\mathrm{J95}}-0.242
\label{transformation_J95_ZW84}
\end{equation}
between the two metallicity scales. Several other metallicity scales are also commonly used. A widely-used scale is that of \citet{Carretta97}. With a new and larger sample, \citet{Carretta09} recalibrated the metallicity scale of \citet{Carretta97} and provide a quadratic relation between their new metallicity scale and the metallicity scale of \citetalias{Zinn84}. Another popular metallicity scale was introduced by \citet{Harris96}, which, according to \citet{Gratton04}, is indistinguishable from \citetalias{Zinn84}.
%
%
\section{Metallicities of the LMC}
\label{FeH_LMC}
\begin{figure}
\centering 
 \includegraphics[height=0.30\textheight,width=0.47\textwidth]{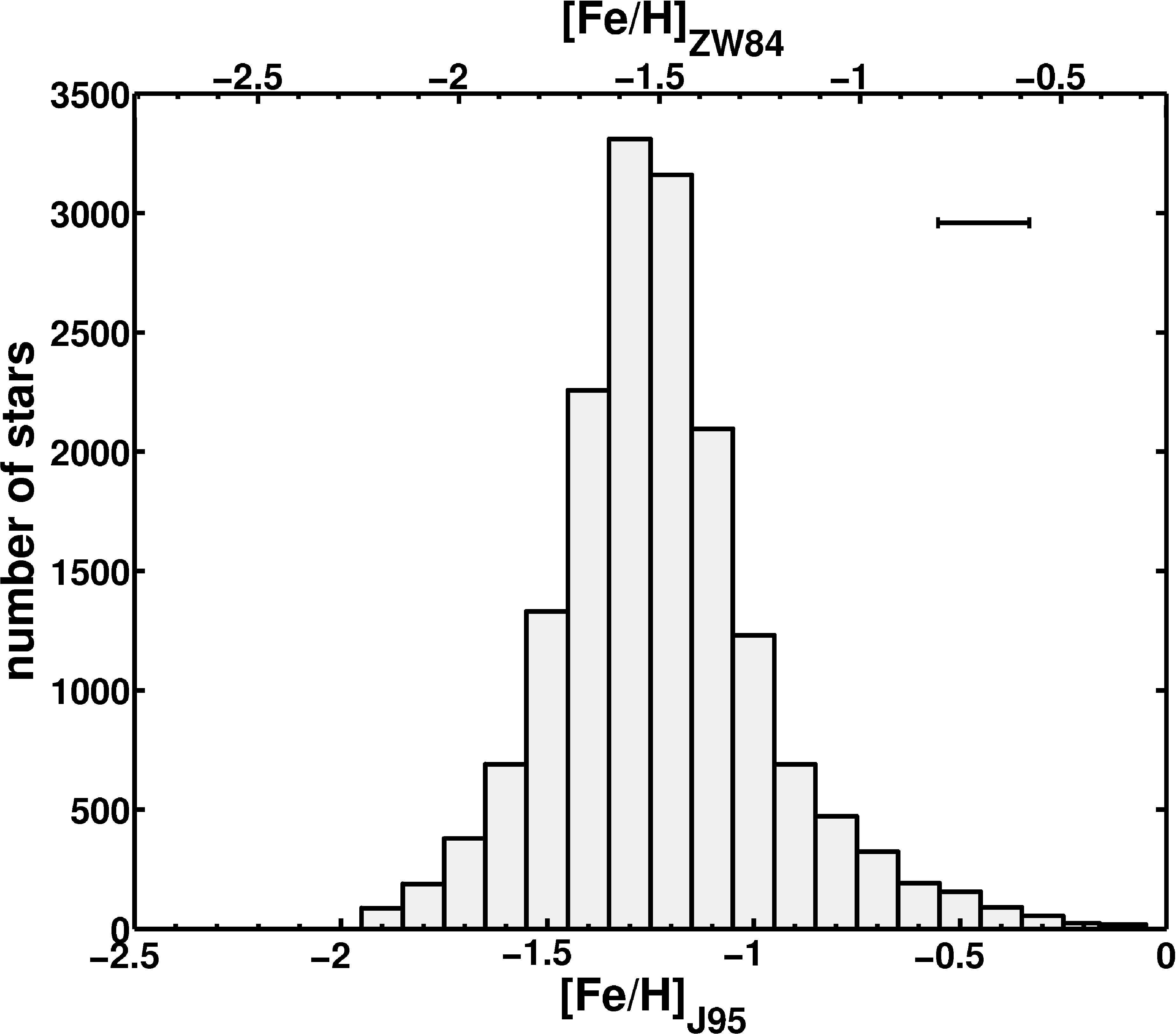}
 \caption{Metallicity distribution of LMC RR\,Lyrae stars, computed based on the Fourier decomposition of their light curves. We find a mean metallicity of $\mathrm{[Fe/H]}_{\mathrm{mean}} = -1.22$\,dex on the scale of \citetalias{Jurcsik95}. The bottom axis shows the scale of \citetalias{Jurcsik95}, while the top axis shows the scale by \citetalias{Zinn84}, for which we find $\mathrm{[Fe/H]}_{\mathrm{mean}} = -1.49$\,dex, using the transformation of \citet{Papadakis00}. The distribution has a standard deviation of 0.26\,dex. We also find a few candidates for very metal-poor stars. The error bar shows the typical uncertainty of the metallicities for individual stars.} 
 \label{MDF_LMC} 
\end{figure}

For all RR\,Lyrae stars of type\,ab in the LMC for which $\phi_{31}$ values are available from the OGLE\,III catalogue, we compute metallicities using Equation\,(\ref{FeH_equation}). This results in a dataset of 16949 stellar metallicities (Figure\,\ref{MDF_LMC}). \\\noindent\hspace*{1em} 
\subsection{Mean metallicities and metallicity distribution function}
Using the \citetalias{Jurcsik95} scale, we find a mean metallicity of $\mathrm{[Fe/H]} = -1.22$\,dex with a standard deviation of 0.26\,dex. This translates to a mean metallicity of $\mathrm{[Fe/H]} = -1.50$\,dex on the \citetalias{Zinn84} metallicity scale. The uncertainties of the Fourier parameters are not mentioned by the OGLE collaboration. Taking the uncertainty of the observed magnitude as the 1\,$\sigma$ standard deviation of its assumed Gaussian error distribution, we randomly vary the observed magnitudes to perform Monte Carlo simulations of the light curve for each individual star and redetermine the Fourier parameters. We find very good agreement between the parameter values obtained by OGLE and via our Monte Carlo approach. The resulting range of parameters allows us to calculate the uncertainty for $\phi_{31}$ and thus of the corresponding metallicity of each star. For the mean uncertainty of the metallicity we find $\sigma_{\textrm{intrinsic}} = 0.10$\,dex. This uncertainty is quadratically added to the systematic uncertainty of $\sigma_{\textrm{sys}} = 0.18$\,dex mentioned by \citet{Smolec05} and results in a total uncertainty of $\sigma_{[Fe/H]} = 0.21$\,dex. 

\begin{figure}
\centering 
 \includegraphics[height=0.30\textheight,width=0.47\textwidth]{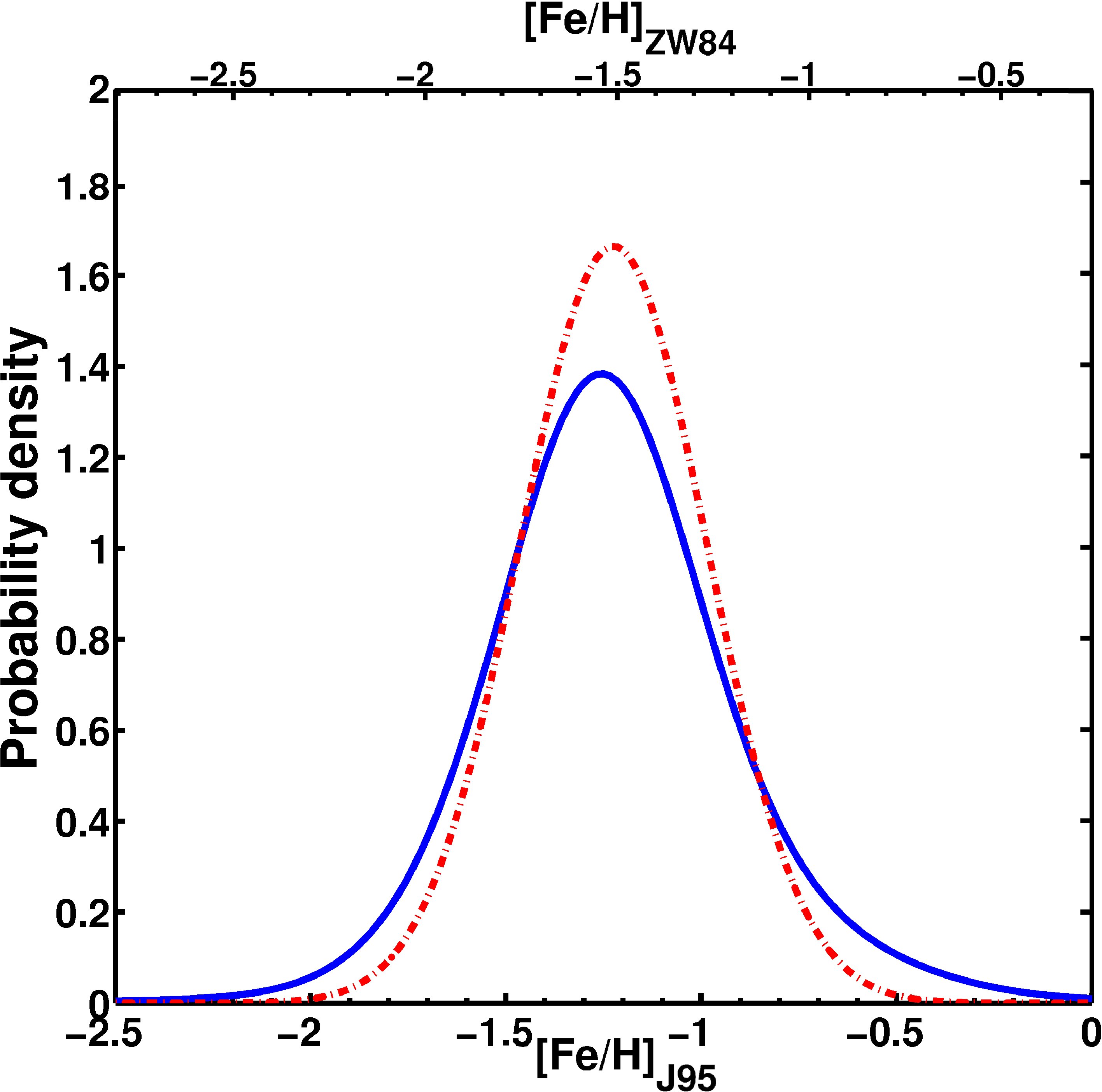}
 \caption{Probability distribution function for the metallicity of LMC RR\,Lyrae stars. The solid blue curve shows the sum of individual stellar metallicities convolved with Gaussian uncertainties, while the red dash-dotted curve is the likelihood distribution. The computation of the metallicities is based on the Fourier decomposition of their light curves. We find a mean metallicity of the likelihood distribution of $\mathrm{[Fe/H]}_{\mathrm{mean}} = -1.23$\,dex on the scale of \citetalias{Jurcsik95} (bottom axis). The top axis shows the \citetalias{Zinn84} scale, with a mean value of $\mathrm{[Fe/H]}_{\mathrm{mean}} = -1.50$\,dex, using the transformation of \citet{Papadakis00}. The MDF has an intrinsic spread of $0.24$\,dex.} 
 \label{MDF_LMC_ML} 
\end{figure}

%
We convolve the individual metallicity values obtained with the derived constant (Gaussian) uncertainty of $0.21$\,dex and sum up the resulting Gaussian distributions to obtain the ``observed'' MDF. The solid blue line in Figure\,\ref{MDF_LMC_ML} shows the MDF normalized to 1 (i.e., a probability distribution function). \\\noindent\hspace*{1em} 
Motivated by the near-Gaussian profile of the observed MDF, we also use a maximum-likelihood approach to constrain the underlying MDF of the LMC RR\,Lyrae population. Using such an approach allows us to determine the {\it{intrinsic}} width of the MDF (assuming a Gaussian profile) while simultaneously accounting for the contribution of the measurement uncertainties to the width of the observed distribution.  The likelihood of our observed data set originating from an underlying Gaussian MDF with its mean, $\mu$, and standard deviation, $\sigma$, is given by the function
\begin{eqnarray}
\mathcal{L}\left(\mu,\sigma \right) = 
& \prod_{i=1}^N \frac{1}{\sqrt{2 \pi \left( \sigma^2 + {\sigma_i}^2 \right)}} \exp \left(-\frac{1}{2} \frac{(x_i - \mu)^2}{ \sigma^2 + {\sigma_i^2 }}\right) 
\label{LH_equation}
\end{eqnarray}
where $N$ is the total number of stars, $x_i$ is the metallicity of star $i$ and $\sigma_i$ is the associated measurement uncertainty. We seek to find the parameters $\mu$ and $\sigma$ that maximize this likelihood function.\\\noindent\hspace*{1em}
The dash-dotted line in Figure\,\ref{MDF_LMC_ML} represents the best fit of $\mu$ and $\sigma$ and reveals the underlying true MDF with a mean metallicity of $\mathrm{[Fe/H]} = -1.23$\,dex and an intrinsic width of the distribution of $0.24$. \\\noindent\hspace*{1em}
The highest metallicity found in our sample is $\mathrm{[Fe/H]}_{max} = 2.0$\,dex, while the lowest value is $\mathrm{[Fe/H]}_{min} = -3.81$\,dex. As explained in Section\,\ref{FeH_method}, we only consider the reliable domain of the calibration and thus exclude 107 stars ($0.6\%$ of the complete sample) at the low-metallicity end and 66 stars ($0.4\%$ of the complete sample) at the high-metallicity end from the dataset when applying this selection. This leaves us with a sample of 16776 RR\,Lyrae stars. However, the mean and the standard deviation of the metallicity of the reduced sample remain unchanged in comparison with the full sample.\\\noindent\hspace*{1em} 
Results on different metallicity scales cannot easily be compared. We transform all metallicities to the scale of \citetalias{Zinn84} using Equation\,(\ref{transformation_J95_ZW84}). This transformation leads to a mean metallicity of $\mathrm{[Fe/H]}_{\mathrm{mean}/\mathrm{ZW84}} = -1.50$\,dex on the \citetalias{Zinn84} scale, which is in excellent agreement with mean values of spectroscopic measurements of RR\,Lyrae metallicities in the LMC \citep[who used the \citetalias{Zinn84} metallicity scale as well]{Gratton04, Borissova04, Borissova06}. \\\noindent\hspace*{1em}
\begin{table}
\caption{}             
\label{table_FeH_literature}      
\centering                          
\begin{tabular}{c c c c}        
\hline\hline                 
study & mean [Fe/H] & $\sigma$ & number of stars\\    
\hline                        
\citet{Gratton04} & $-1.48$ & 0.03 & 98\\
\citet{Borissova04} & $-1.46$ & 0.09 & 74\\
\citet{Borissova06} & $-1.53$ & 0.02 & 78\\
\hline                                   
\end{tabular}
\end{table}
\subsection{Spatial variations in metallicity}
\begin{figure}
\centering 
 \includegraphics[height=0.30\textheight,width=0.47\textwidth]{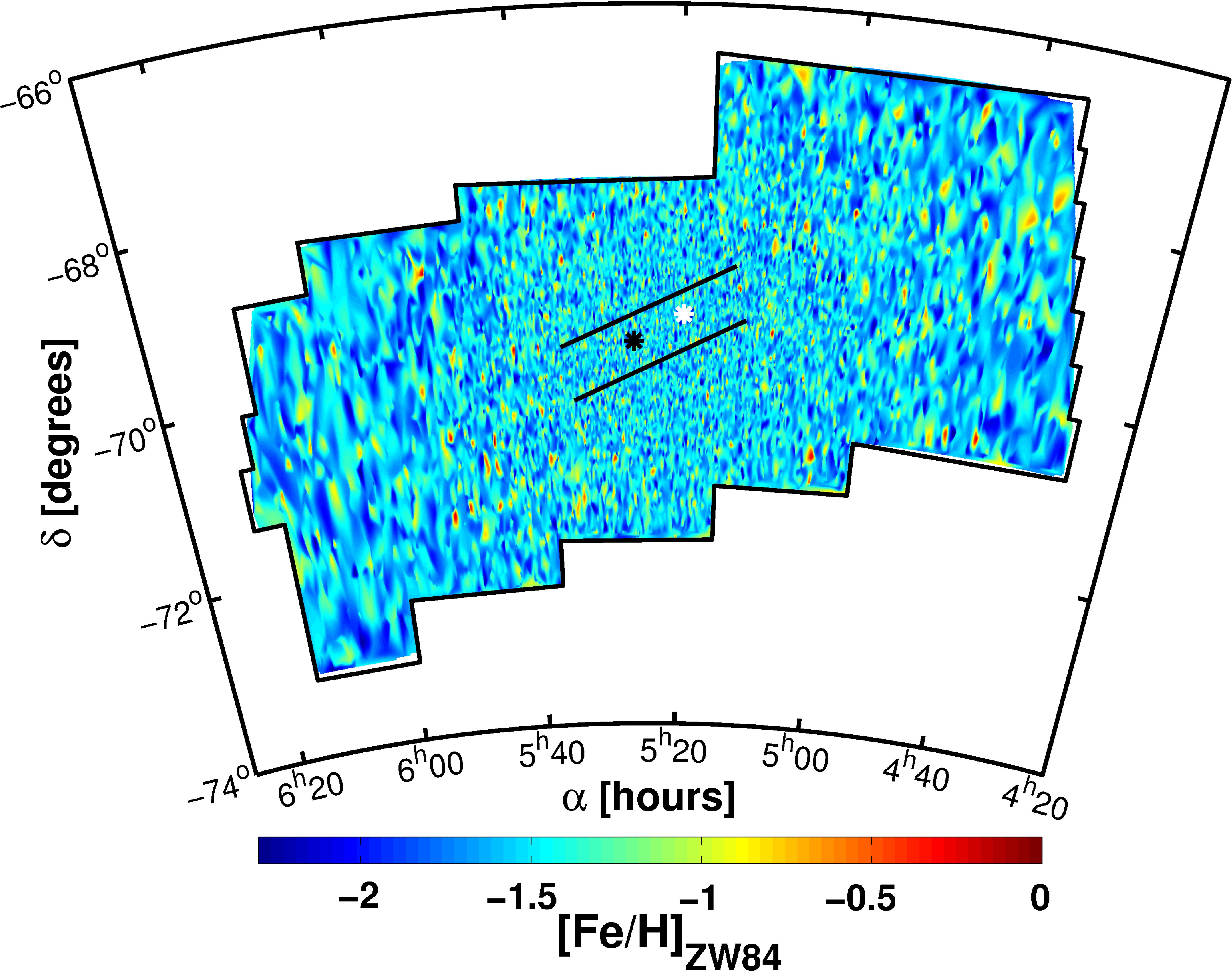}
\caption{Distribution of metallicity in the LMC. Metal-poor regions are coded in blue, metal-rich regions are red. Overall the contour plot is smooth and no gradient or distinct feature is visible. For the contour plot, only stars with $-2.3 < \mathrm{[Fe/H]}_{\mathrm{ZW84}} < 0.0$ are taken into account. The black lines in the middle represent the bar. While the white asterisk marks the optical center of the LMC found by \citet{Vaucouleurs72}, the black asterisk shows the center of the RR\,Lyrae stars as found by \citet{Haschke11_LMC}.} 
 \label{metallicity_distribution_contour_LMC} 
\end{figure}

\begin{figure}
 \includegraphics[width=0.47\textwidth]{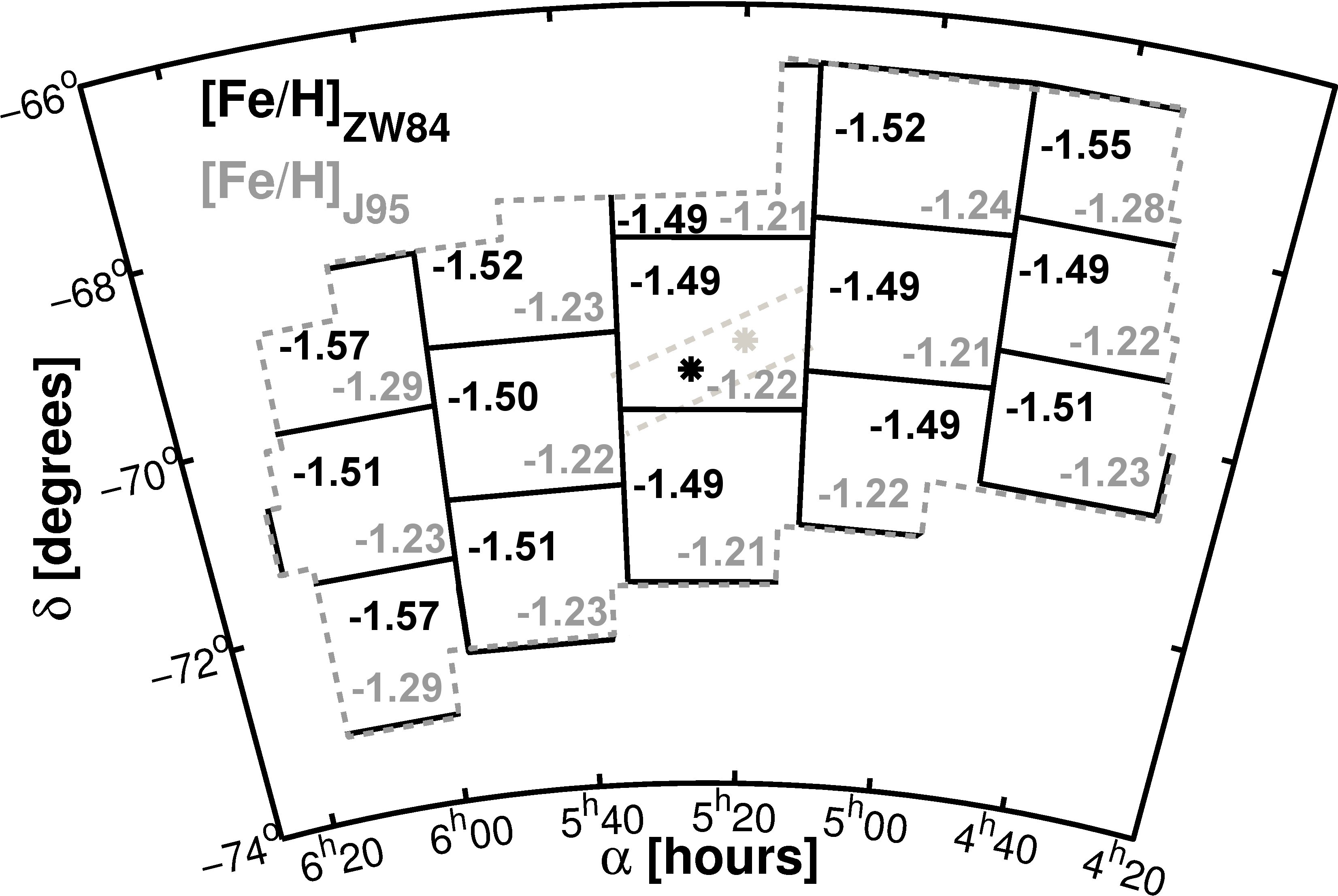}
 \caption{The variance of metallicity between the inner- and outermost fields is tested by dividing the LMC field into 15 similarly sized fields. The mean metallicity values on the \citetalias{Zinn84}-scale are shown in black, grey values represent the scale of \citet{Jurcsik95}. We see a small decline of the metallicities in the outermost regions sampled by OGLE\,III. The boundaries of the bar are illustrated with the dashed grey lines, while the black asterisk marks the center as found by RR\,Lyrae stars and the grey asterisk the optical center of the LMC.} 
 \label{metallicity_distribution_LMC} 
\end{figure}

Figure\,\ref{metallicity_distribution_contour_LMC} shows the spatial distribution of the metallicities. Only stars with a metallicity from the reliable domain of the \citet{Smolec05} calibration are used in this figure. Individual stars are used for the contour plot, to which a grid with constant boxwidth of $1.5' \times 1.5'$ was then applied. Figure\,\ref{metallicity_distribution_contour_LMC} shows the median metallicity values of these boxes.\\\noindent\hspace*{1em}
In general, a smooth distribution with small-scale variations, but without any obvious gradients or distinct features, is visible. Towards the center of the LMC, the density of RR\,Lyrae stars increases. Differences in metallicity are therefore observed on smaller scales in the central parts than in the outskirts. This leads to the higher resolution in the center of the figure. The stars that are excluded due to their metallicity being above or below the adopted validity range are distributed randomly across the whole field of OGLE\,III.\\\noindent\hspace*{1em}
\begin{figure}
 \includegraphics[width=0.47\textwidth]{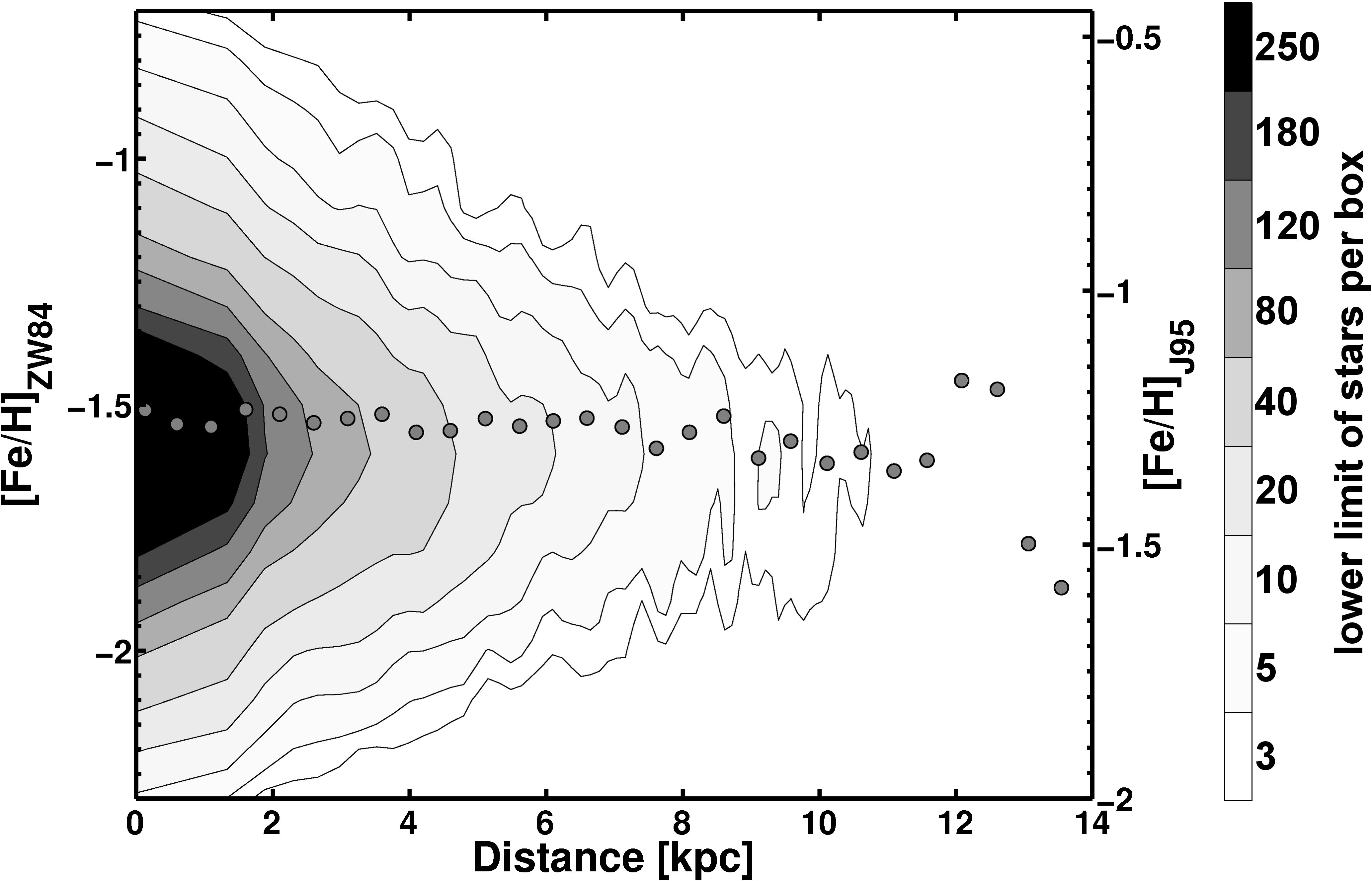}
 \caption{The contours show the total number of RR\,Lyrae stars of a given metallicity as a function of the projected distance relative to the center of the LMC as found from RR\,Lyrae stars. The mean metallicity does not change, within the errors, with increasing distance from the center. The grey dots represent the median metallicity in bins of 0.5\,kpc.} 
 \label{metallicity_distance_fit_LMC} 
\end{figure}
Figure\,\ref{metallicity_distance_fit_LMC} examines the mean metallicity as a function of the distance from the center of the RR\,Lyrae stars of the LMC at $\alpha = 5^h26^m$ and $\delta = -69^\circ75'$ as found by \citet{Haschke11_LMC}. We investigate the metallicity and distance distributions in bins of 0.1\,dex and rings of varying radius. The radii are chosen such that all annuli have the same area. The density of stars in each box is determined. The contour plot is smoothed with a Gaussian filter using $3 \times 3$ boxes and a width of one box. Additionally we calculate the median metallicity of each distance bin of 0.5\,kpc width, this \textquotedblleft running median$\textquotedblright$ is represented by the grey dots in Figure\,\ref{metallicity_distance_fit_LMC}. \\\noindent\hspace*{1em}
To determine the gradient of the metallicity, the median metallicity is computed in bins of 0.1\,kpc and fitted using a first-order polynomial. The resulting gradient for the whole distribution of $-0.03 \pm 0.07\,$dex kpc$^{-1}$ is consistent with zero. If we take only the innermost 8\,kpc from the center of the RR\,Lyrae stars into account, a gradient of $-0.010 \pm 0.014\,$dex kpc$^{-1}$ is found. The median metallicity seems to be constant in the inner parts of the LMC and dropping slightly towards the outskirts.  \\\noindent\hspace*{1em}
To test for non-radial gradients, we divide the OGLE\,III field into 15 similarly-sized fields. As shown in Figure\,\ref{metallicity_distribution_LMC}, the mean metallicity of these fields drops in accordance with the very small metallicity gradient obtained above. The outer fields always show slightly lower metallicities than the innermost fields. The median values are nearly identical to the mean values.\\\noindent\hspace*{1em}
\citetalias{Feast10} determined the metallicity gradient of the old population using OGLE\,III RR\,Lyrae stars. These authors calculated photometric metallicities with two sets of equations, both with the period as the only free parameter. The uncertainty of this method is therefore higher ($\sigma = 0.46$\,dex) than with the method of \citet{Jurcsik96} ($\sigma = 0.16$\,dex). We recalculate our metallicities using the equations utilized by \citetalias{Feast10} and compare them with our estimates (see Table\,\ref{table_FeH_comparison_Feast}). Equation\,(2) of \citetalias{Feast10} and our estimates transformed to the \citetalias{Zinn84} metallicity scale lead to similar results when averaging over all metallicities. For subsets of the sample, however, we find considerable differences. We divide the sample into subsets where all stars are more metal rich than our average or more metal poor. The more metal-rich stars have in general higher metallicities in our Fourier-based estimates.  The metallicities below the mean are in general lower in our calculations than the values computed with Equation\,(2) by \citetalias{Feast10}. When only considering stars at the extremes of our metallicity estimates this trend becomes even stronger (Table\,\ref{table_FeH_comparison_Feast}). However, the spatial distribution of the RR\,Lyrae stars in each metallicity bin is very similar to the overall distribution.  \\\noindent\hspace*{1em}
\begin{table*}
\caption{Differences between metallicity estimates by \citetalias{Feast10} and by us for different ranges of metallicity. The first column gives the metallicity scale we adopt, the second column the equation used by \citetalias{Feast10} in order to derive the metallicity, the third column the metallicity range used for the comparison is shown, and the fourth column shows the difference found by calculating $\Delta_{HF} = \mathrm{[Fe/H]}_{\mathrm{HGDJ11}} - \mathrm{[Fe/H]}_{\mathrm{FAW10}}$.  }             
\label{table_FeH_comparison_Feast}      
\centering                          
\begin{tabular}{c c c r}        
\hline\hline                 
$\mathrm{[Fe/H]}_{\mathrm{HGDJ11}}$ scale  & $\mathrm{[Fe/H]}_{\mathrm{FAW10}}$  equation & $\mathrm{[Fe/H]}_{\mathrm{J95}}$ range & $\Delta_{HF}$ \\    
\hline                        
J95 & $\mathrm{[Fe/H]} = -5.62\log P - 2.81$ & $-2.00 < \mathrm{[Fe/H]} < +0.20$ & 0.17 \\ 
J95 & $\mathrm{[Fe/H]} = -7.82\log P - 3.43$ & $-2.00 < \mathrm{[Fe/H]} < +0.20$ & 0.25 \\ 
ZW84 & $\mathrm{[Fe/H]} = -5.62\log P - 2.81$ & $-2.00 < \mathrm{[Fe/H]} < +0.20$ & $-0.11$ \\
ZW84 & $\mathrm{[Fe/H]} = -7.82\log P - 3.43$ & $-2.00 < \mathrm{[Fe/H]} < +0.20$ & $-0.03$ \\
ZW84 & $\mathrm{[Fe/H]} = -7.82\log P - 3.43$ & $-1.23 < \mathrm{[Fe/H]} < +0.20$ & $0.23$ \\
ZW84 & $\mathrm{[Fe/H]} = -7.82\log P - 3.43$ & $-2.00 < \mathrm{[Fe/H]} < -1.23$ & $-0.18$ \\
ZW84 & $\mathrm{[Fe/H]} = -7.82\log P - 3.43$ & $-0.50 < \mathrm{[Fe/H]} < +0.00$ & $0.52$ \\
ZW84 & $\mathrm{[Fe/H]} = -7.82\log P - 3.43$ & $-1.00 < \mathrm{[Fe/H]} < -0.50$ & $0.41$ \\
ZW84 & $\mathrm{[Fe/H]} = -7.82\log P - 3.43$ & $-1.50 < \mathrm{[Fe/H]} < -1.00$ & $-0.05$ \\
ZW84 & $\mathrm{[Fe/H]} = -7.82\log P - 3.43$ & $-2.00 < \mathrm{[Fe/H]} < -1.50$ & $-0.17$ \\
\hline                                   
\end{tabular}
\end{table*}
\citetalias{Feast10} conclude that there is a very small gradient of $-0.0104 \pm 0.0021\,$dex kpc$^{-1}$ for the RR\,Lyrae stars of the innermost 6\,kpc of the LMC. This result is in very good agreement with the gradient found in our study. Our error estimate is a bit more conservative, as we take the intrinsic uncertainties of the metallicity into account. Within the errors, we are not able to distinguish between no gradient and a very small gradient. \\\noindent\hspace*{1em}
The result of essentially no metallicity gradient confirms the findings of \citet{Grocholski06, Cole09} and \citet{Sharma10}, who used either star clusters with ages between 1 and 10\,Gyr or red giants to determine the metallicity gradient of the LMC. A small gradient of $-0.047 \pm 0.003$\,dex kpc$^{-1}$ was obtained by \citet{Cioni09} using AGB stars, which span an age range of $3-10$\,Gyr. Based on spectroscopy of red giants, \citet{Carrera11} find that the metallicity of these stars in the inner 6\,kpc of the LMC is roughly constant, but then decreases with increasing radii. This result is supported by our findings with the (on-average) older RR\,Lyrae stars.
%
%
\section{Metallicities of the SMC}
\label{FeH_SMC}
\begin{figure}
\centering 
 \includegraphics[height=0.30\textheight,width=0.47\textwidth]{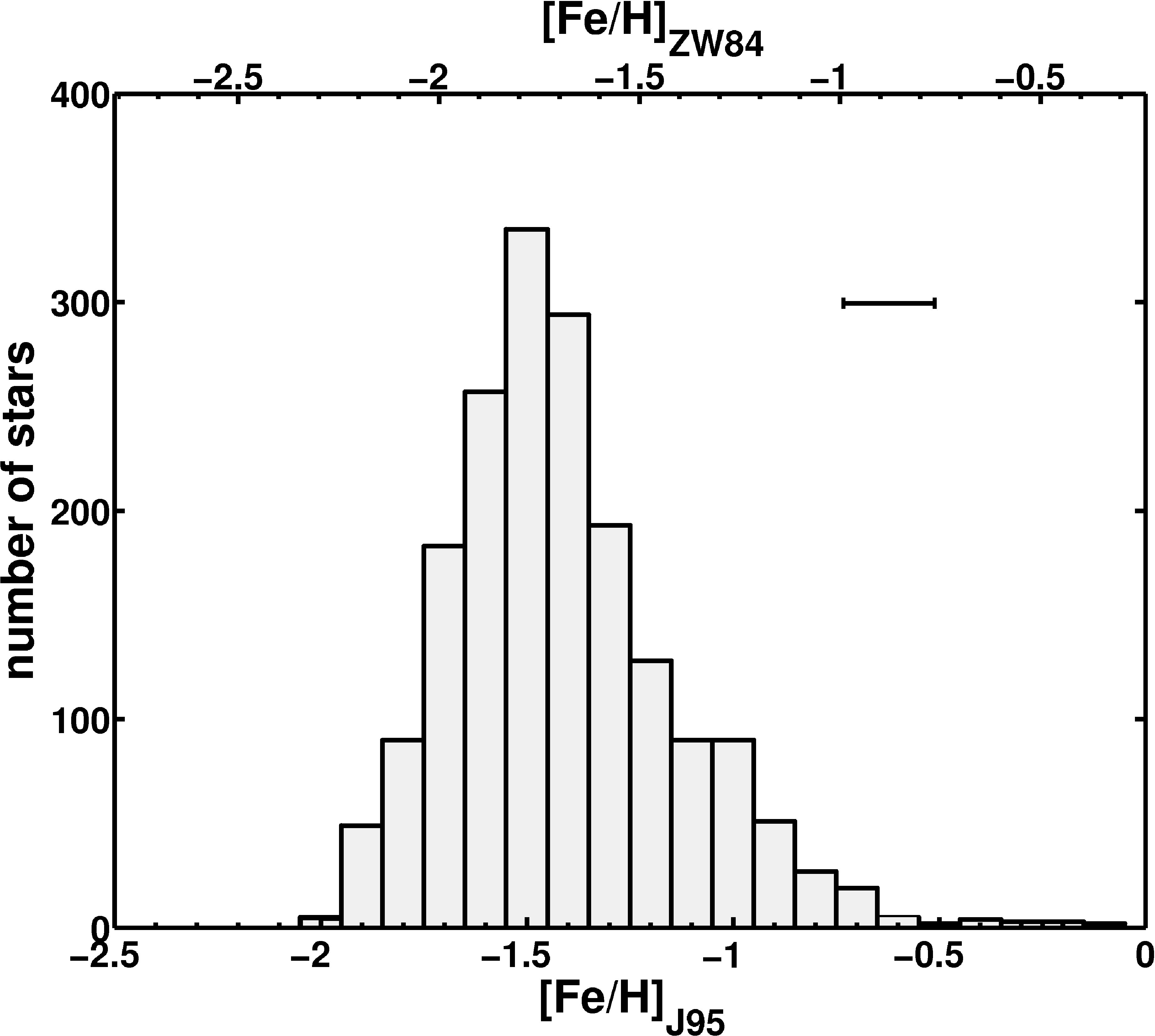}
 \caption{The metallicity distribution for the old population of the SMC. For 1864 RR\,Lyrae stars [Fe/H] is determined and a mean metallicity of $\mathrm{[Fe/H]}_{\mathrm{mean}} = -1.42$\,dex on the scale of \citetalias{Jurcsik95} is obtained (bottom axis), while on the scale of \citetalias{Zinn84} we find $\mathrm{[Fe/H]}_{\mathrm{mean}} = -1.70$\,dex (top axis). The standard deviation of the distribution is 0.33\,dex. The error bar represents the typical uncertainty of the metallicities for individual stars.} 
 \label{MDF_SMC} 
\end{figure}

\begin{figure}
\centering 
 \includegraphics[height=0.30\textheight,width=0.47\textwidth]{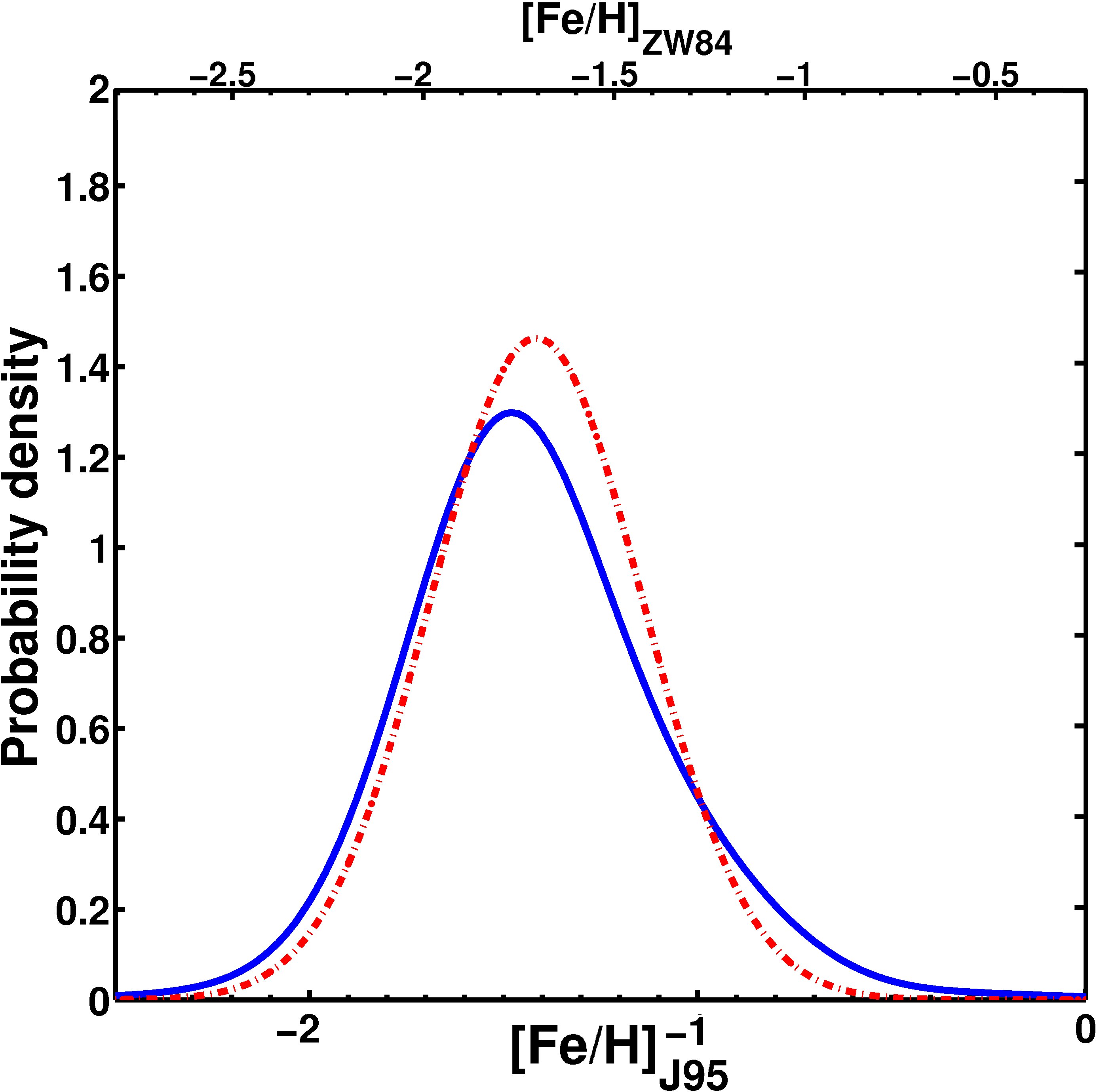}
 \caption{Metallicity probability distribution functions for the SMC. The solid blue curve represents the measured metallicities convolved with Gaussian uncertainties. A likelihood distribution is determined from the solid curve and given by the red dash-dotted line. We use 1831 RR\,Lyrae stars and find a mean metallicity of $\mathrm{[Fe/H]}_{\mathrm{mean}} = -1.42$\,dex on the scale by \citetalias{Jurcsik95} (bottom axis). $\mathrm{[Fe/H]}_{\mathrm{mean}} = -1.70$\,dex is obtained for the scale of \citetalias{Zinn84} (top axis). The intrinsic spread of the distribution is 0.27\,dex.} 
 \label{MDF_SMC_ML} 
\end{figure}

\begin{figure}
\centering 
 \includegraphics[height=0.30\textheight,width=0.47\textwidth]{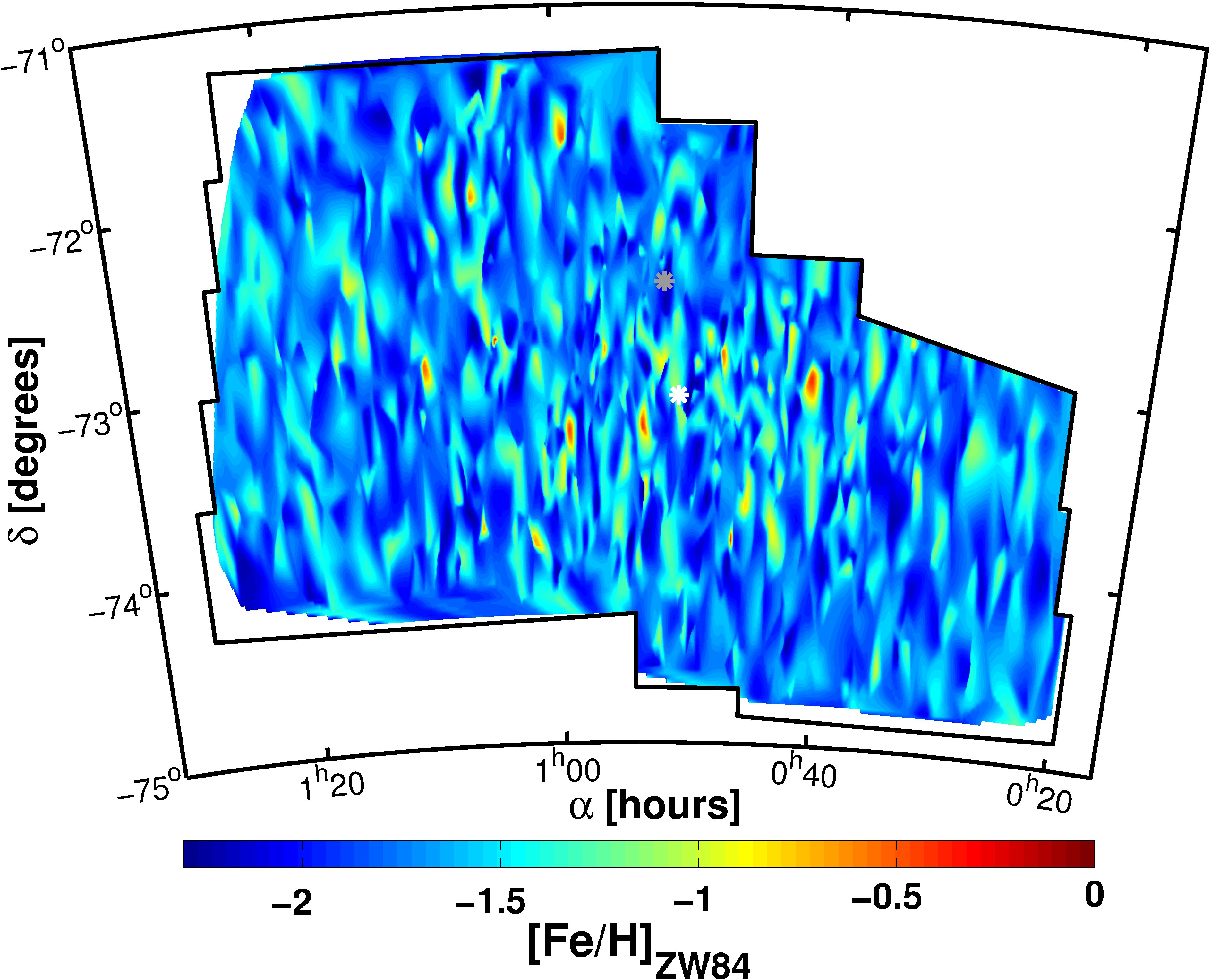}
 \caption{Distribution of metallicity in the SMC. Metal-poor regions are coded in blue, metal-rich regions are red. Overall the contour plot is smooth and no gradient or distinct feature is visible. For the contour plot only stars with $-2.0 < \mathrm{[Fe/H]} < +0.2$ are taken into account. The white asterisk represents the center found by \citet{Gonidakis09}, the grey asterisk the center of \citet{Piatek08}.} 
 \label{metallicity_distribution_contour_SMC} 
\end{figure}

\begin{figure}
 \includegraphics[width=0.47\textwidth]{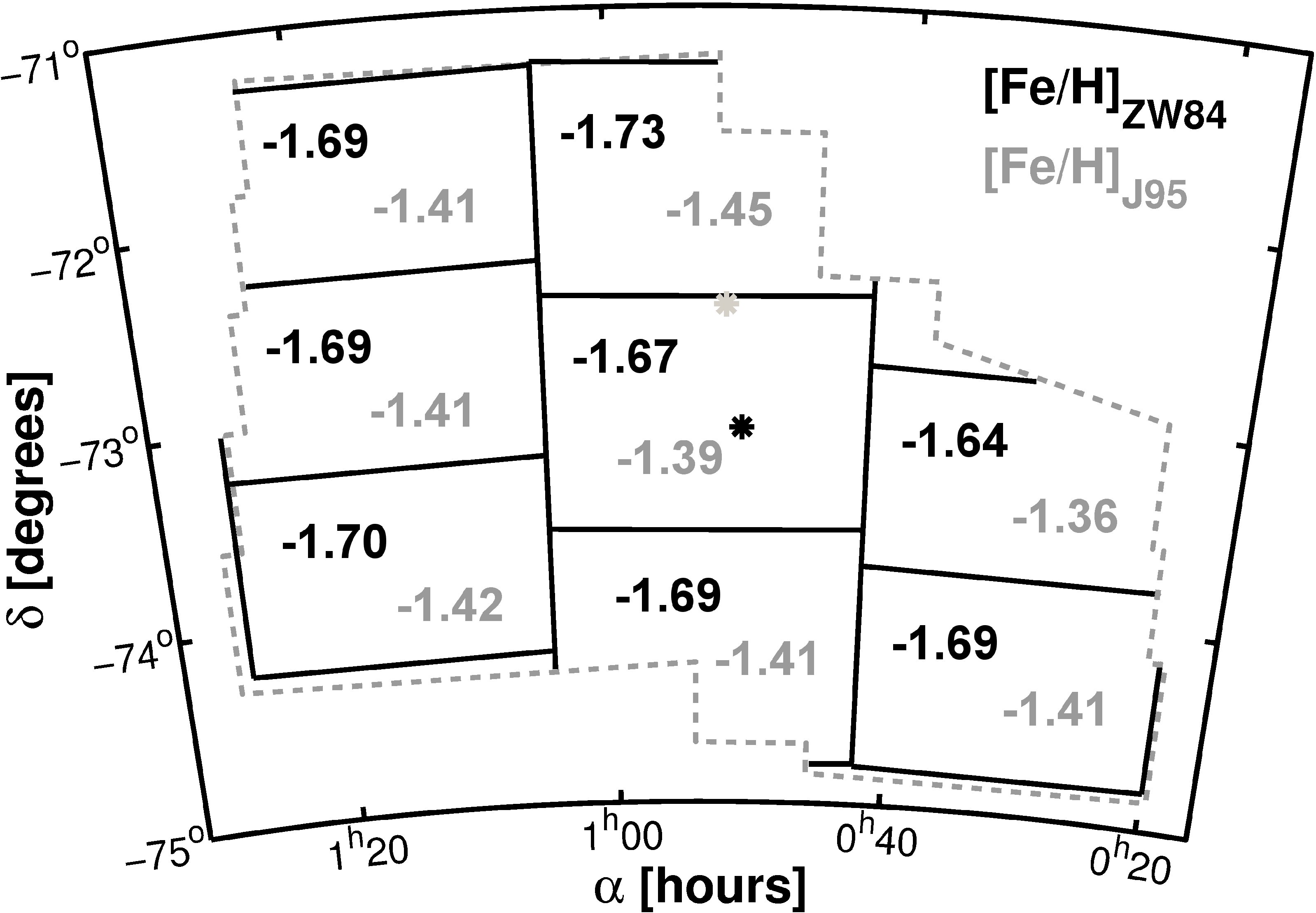}
 \caption{The SMC field is divided into nine similarly sized fields and the mean metallicity for each field is obtained. The mean values on the \citetalias{Jurcsik95} and \citetalias{Zinn84} scales are shown in grey and black, respectively. The values of the different fields are very similar to each other. The black and grey asterisk indicates the center of the SMC as found by \citet{Gonidakis09} and \citet{Piatek08}, respectively. } 
 \label{metallicity_distribution_SMC} 
\end{figure}

In the SMC, Fourier parameters were determined for 1864 RR\,Lyrae stars of type\,ab by the OGLE\,III collaboration \citep{Soszynski10b}. The resulting metallicity distribution is shown in Figure\,\ref{MDF_SMC}. \\\noindent\hspace*{1em} 
Following the same maximum-likelihood approach as described for the LMC RR\,Lyrae, we find the underlying metallicity distribution function of SMC RR\,Lyrae to have a mean value of $\mathrm{[Fe/H]} = -1.42$\,dex on the scale of \citetalias{Jurcsik95} with a standard deviation of $0.27$\,dex as shown in Figure\,\ref{MDF_SMC_ML}. As explained in Section\,\ref{FeH_method}, we choose the metallicity regime between $-2.0 < \mathrm{[Fe/H]} < +0.2$ on the scale of \citetalias{Jurcsik95} as the reliable domain of the calibration by \citet{Smolec05}. All values outside of this calibration range are rejected. 27 stars with metallicities below $\mathrm{[Fe/H]} = -2.0$\,dex are found, while the lowest value is measured to be $\mathrm{[Fe/H]}_{min} = -3.3$\,dex. Only 6 RR\,Lyrae with metallicities above $\mathrm{[Fe/H]} = +0.2$\,dex are present and a maximum of $\mathrm{[Fe/H]}_{max} = 1.7$\,dex is found. Eliminating these stars, which are randomly distributed in space, leaves us with a sample of 1831 RR\,Lyrae stars with presumably reliable metallicities. \\\noindent\hspace*{1em} 
The resulting mean metallicity of the old population of the SMC is $\mathrm{[Fe/H]}_{\mathrm{mean}} = -1.42$\,dex on the scale of \citetalias{Jurcsik95} and $\mathrm{[Fe/H]}_{\mathrm{mean}} = -1.70$\,dex on the \citetalias{Zinn84} metallicity scale. So far no spectroscopic measurements of RR\,Lyrae field stars have been published. We therefore compare our result with the metallicity of NGC\,121, the only globular cluster of the SMC in the age range bracketed by our RR\,Lyrae stars. \citet{Glatt08a} found an age of 10.5\,Gyr for NGC\,121 using main-sequence fitting with Dartmouth \citep{Dotter07} isochrones. \citet{DaCosta98} used the Ca\,II triplet to determine the mean metallicity of NGC\,121 and found $\mathrm{[Fe/H]} = -1.46 \pm 0.10$ on the scale of \citetalias{Zinn84}, in very good agreement with the results we obtain for the old population of the SMC. \\\noindent\hspace*{1em}
Recently \citet{Kapakos11} used a sample of 100\,RR\,Lyrae stars from the OGLE\,III catalogue to determine photometric metallicities using the relation by \citet{Jurcsik96}. They decomposed the $V$ band light curves to determine the Fourier parameters of each star. On the metallicity scale of \citetalias{Jurcsik95} they found a mean metallicity of $\mathrm{[Fe/H]}_{\mathrm{mean}} = -1.51 \pm 0.41$\,dex. This is again in very good agreement with the metallicities found with our sample. \\\noindent\hspace*{1em}
We use our metallicity estimates on the scale of \citetalias{Jurcsik95} for the individual RR\,Lyrae stars to plot a contour map of the metallicity distribution of the old population (Figure\,\ref{metallicity_distribution_contour_SMC}). The overall distribution of the metallicity is quite smooth and no distinct features are visible. The black lines indicate the borders of the OGLE\,III field. In the white regions within this polygon no RR\,Lyrae stars were detected and therefore no contours are drawn. \\\noindent\hspace*{1em}
\begin{figure}
 \includegraphics[width=0.47\textwidth]{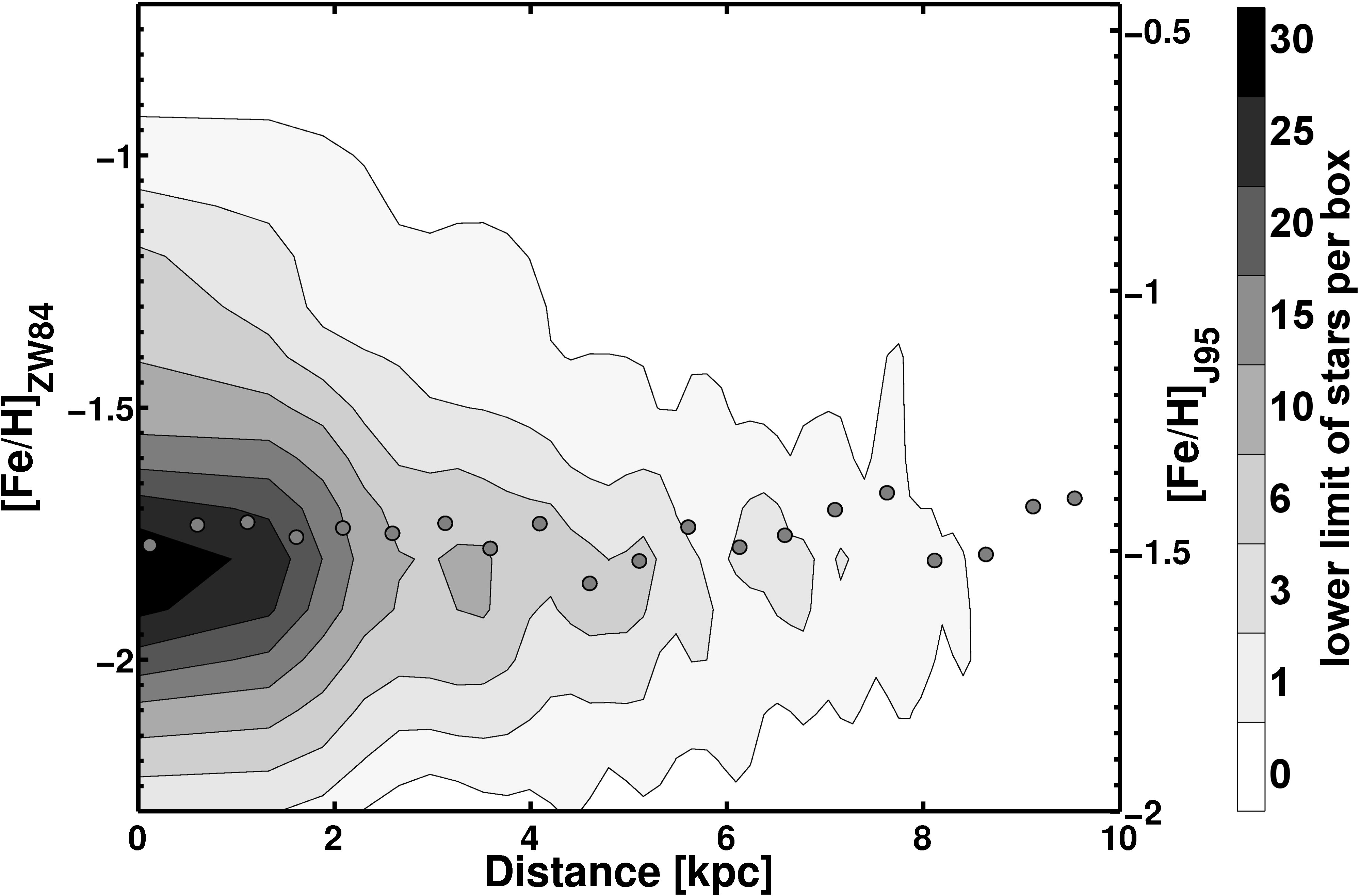}
 \caption{The total number of stars in boxes of metallicity versus the projected distance from the center as determined by \citet{Gonidakis09} (white star in Figure\,\ref{metallicity_distribution_contour_SMC}). The grey dots represent the median metallicity of small distance bins of 0.5\,kpc. Within the uncertainties no radial gradient of the metallicity is found.} 
 \label{metallicity_distance_fit_SMC} 
\end{figure}
The metallicity gradient is measured with respect to the center of the SMC determined by K-, M- and faint carbon stars \citep[$\alpha = 0^h51^m$ and $\delta = -73^\circ7'$,][]{Gonidakis09} This center is indicated in Figure\,\ref{metallicity_distribution_contour_SMC} with a white asterisk. The spatial density distribution of RR\,Lyrae stars is bimodal \citep{Haschke11_SMC} and the determination of a single center is therefore rather difficult. We have chosen to adopt the values by \citet{Gonidakis09}, because they represent the mean center of the RR\,Lyrae distribution quite well. The stellar density is evaluated in boxes of 0.1\,dex with a changing radius to keep the total area of each annulus constant. The resulting contours are smoothed with a Gaussian kernel and shown in Figure\,\ref{metallicity_distance_fit_SMC}. \\\noindent\hspace*{1em}
In order to investigate the presence of a gradient we fit a first-order polynomial to the running median of the SMC. We find a gradient of $0.00 \pm 0.06$\,dex kpc$^{-1}$. We subdivide the OGLE\,III field of the SMC into 9 fields of similar area and determine the mean metallicity in each (Figure\,\ref{metallicity_distribution_SMC}). We note that the mean is very similar to the median value. The metallicity stays basically constant, as expected from the essentially zero gradient. \\\noindent\hspace*{1em}
\citet{Piatek08} used HST proper motion data to determine the kinematical center of the SMC. They found $\alpha = 0^h52^m8^s$ and $\delta = -72^\circ30'$, which is shown by the grey star in Figure\,\ref{metallicity_distribution_contour_SMC}. We do not find pronounced radial trends in the metallicity when using this center of the SMC either. \\\noindent\hspace*{1em}
Metallicity gradients of the SMC have been studied using star clusters and red giants. \citet{DaCosta98} used spectroscopic measurements of giants and star clusters to search for a possible metallicity gradient and did not find any evidence. \citet{Glatt08b} studied six intermediate age clusters and found a spread of $0.6$\,dex between the most metal-poor and most metal-rich clusters. The spectroscopic metallicity estimates lead to the conclusion that no smooth, monotonic age metallicity relation is present, but that instead there is a spread in metallicity at any given age \citep{Kayser07, Glatt08b}. Using spectroscopic data of red giants \citet{Parisi09} did not find a gradient, while \citet{Carrera08b} suggest that one exists using a different sample of spectroscopic observations of red giant field stars. \citet{Carrera08b} argue that this gradient is the result of different ages of the red giants at different locations of the SMC. They conclude that further star formation has taken place in the inner parts and led to the metallicity gradient among these intermediate-age stars with ages between 3 and 10\,Gyr. However this would not have affected our old RR\,Lyrae stars. \\\noindent\hspace*{1em}
%
%
\section{Comparison of spectroscopic and photometric metallicities}
\label{FeH_spec_phot}
\begin{figure}
\centering 
 \includegraphics[width=0.47\textwidth]{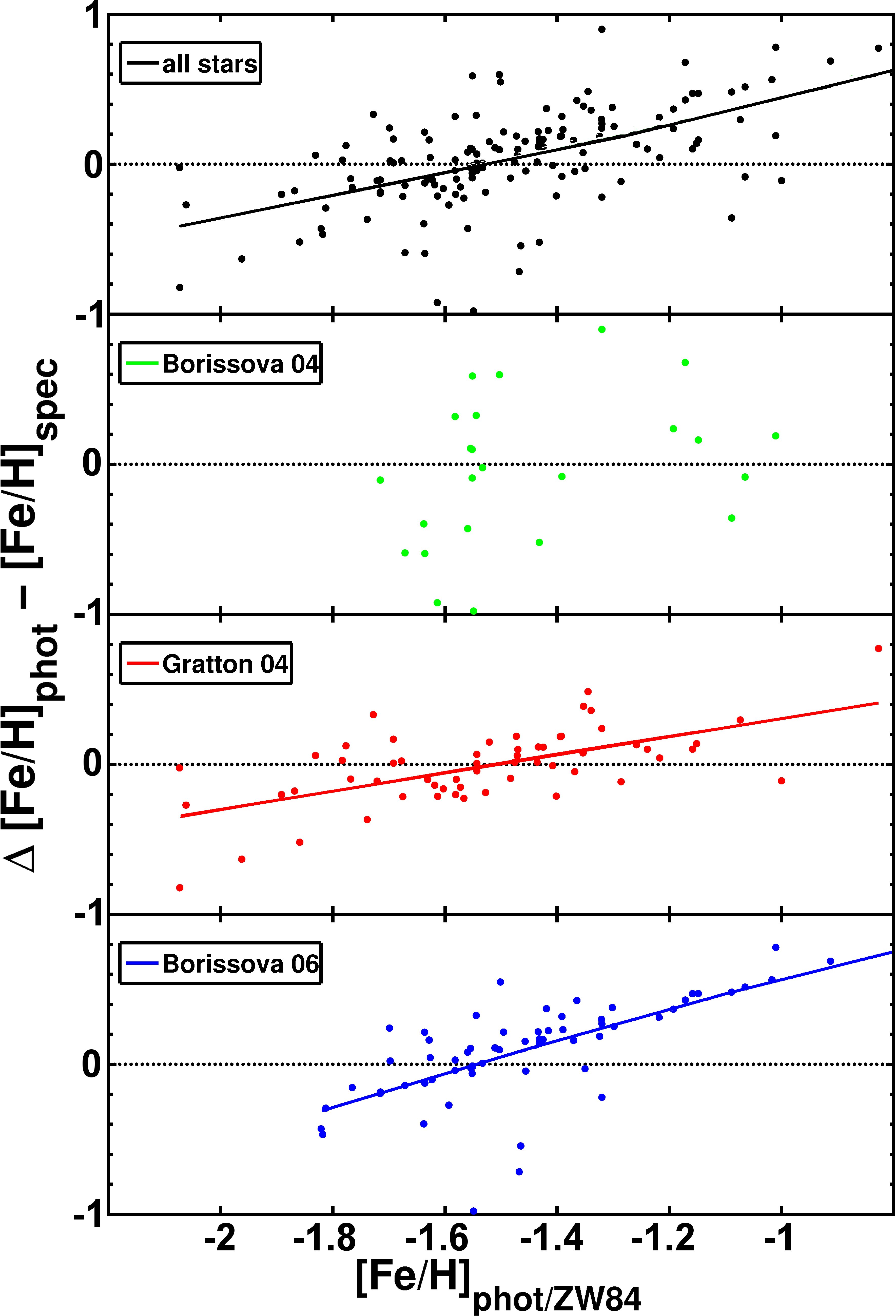}
 \caption{Comparison of our photometric metallicities with spectroscopic metallicities from the literature for LMC RR\,Lyrae stars. All metallicities are on the \citetalias{Zinn84} scale. Calculating $\Delta\mathrm{[Fe/H]} = \mathrm{[Fe/H]}_{phot} - \mathrm{[Fe/H]}_{spec}$ we find a median difference of $0.05$\,dex for stars with photometric metallicities and their counterparts in the samples of \citet{Gratton04} and \citet{Borissova04, Borissova06}. The trends in each sample, except for \citet{Borissova04}, are fitted with a linear polynomial.}
 \label{comparison_phot_deltaphotspec} 
\end{figure}

For some RR\,Lyrae stars located in the bar of the LMC, spectroscopic metallicities were determined in earlier studies. 149 LMC RR\,Lyrae with spectroscopic metallicities from \citet{Gratton04} and \citet{Borissova04, Borissova06} also have OGLE light curves and are thus part of our sample. Figure\,\ref{comparison_phot_deltaphotspec} shows the photometric metallicities we determine with the Fourier decomposition method versus the difference of photometric minus spectroscopic metallicities from the literature. \\\noindent\hspace*{1em} 
The spectroscopic measurements of \citet{Borissova04} are based on the method of \citet{Layden94}, who used the comparison of the equivalent width of Ca\,II\,K line with hydrogen lines. The resulting metallicity is on the scale of \citetalias{Zinn84}. \citet{Borissova06} and \citet{Gratton04} used the method of line indices \citep{Preston59} as outlined in \citet{Gratton04}. This method is linked to the metallicity scale of \citet{Harris96}, which has a very small offset of $+0.02 \pm 0.01$\,dex compared to \citetalias{Zinn84} \citep{Gratton04}. Due to the nearly negligible offset we do not adjust the metallicity values when applying the comparison to the photometric metallicities. \\\noindent\hspace*{1em} 
In Figure\,\ref{comparison_phot_deltaphotspec} we see that there appears to be a systematic trend in the deviation between spectroscopic and photometric measurements such that for low metallicities (lower than $\sim -1.8$\,dex on the \citetalias{Zinn84} scale) the spectroscopic metallicities are typically higher, while for photometric metallicities higher than $\sim -1.3$\,dex the spectroscopic values are on average increasingly lower. These trends are fitted with linear polynomials (Figure\,\ref{comparison_phot_deltaphotspec} and Table\,\ref{table_phot_spec_FeH}). The median deviation, when subtracting the spectroscopic metallicity from the photometric metallicity, is $+0.05$\,dex, while we find a standard deviation of 0.35\,dex. However, the spectroscopic values in the three studies cited above show very different characteristics, as presented in Table\,\ref{table_phot_spec_FeH}. For the 58\,RR\,Lyrae from the sample by \citet{Gratton04} and the 28\,stars in common with \citet{Borissova04} we find a median deviation of 0.01\,dex and $-0.05$\,dex with a standard deviation of 0.25\,dex and 0.49\,dex, respectively, whereas we find a median difference of 0.15\,dex and a standard deviation of 0.35\,dex for the 66\,RR\,Lyrae taken from \citet{Borissova06}. They quote an uncertainty range of 0.20\,dex for their spectroscopic metallicities.   \\\noindent\hspace*{1em} 
\begin{table*}
\caption{Differences between the photometric and spectroscopic metallicities for a sample of 149 RR\,Lyrae stars. We calculate the mean, median and standard deviation for the whole set of stars and for every study. To test for systematics we fit the distribution with a linear least-square fit.}             
\label{table_phot_spec_FeH}      
\centering                          
\begin{tabular}{r r r r r r}        
\hline\hline                 
& all stars & \citet{Gratton04} & \citet{Borissova04} & \citet{Borissova06} \\
\hline                        
mean & 0.037 & $-0.008$ & $-0.041$ & 0.108 \\
median & 0.051 & 0.011 & $-0.052$ & 0.145 \\
std & 0.346 & 0.254 & 0.492 & 0.347 \\
fit & $0.8160\mathrm{[Fe/H]} + 1.2427$ & $0.6181\mathrm{[Fe/H]} + 0.9388$ & --- & $1.0272\mathrm{[Fe/H]} + 1.5884$ \\
\hline                                   
\end{tabular}
\end{table*}
In \citet{Kunder08} the same spectroscopic metallicities as used here are compared with the photometric data of the $V$ band lightcurves of the LMC RR\,Lyrae stars from the MACHO survey. They do not find any evidence for a systematic discrepancy between the spectroscopic values and the photometric metallicities calculated with the relation by \citet[our Equation\,\ref{Jurcsik96_FeH}]{Jurcsik96}. This leads to the conclusion that the systematic difference in our study is introduced by using the relation by \citet{Smolec05}. However, most of the stars are located in the photometric metallicity regime where the differences to the spectroscopic measurements are quite small. \\\noindent\hspace*{1em} 
We also searched the spectroscopically studied, presumably old, star clusters of \citet{Colucci11, Grocholski06} and \citet{Mucciarelli10} for RR\,Lyrae stars detected by OGLE. We do not find any coincidences of RR\,Lyrae stars in our sample within the clusters' diameter.
%

%

\section{Metallicities from transformed $\phi_{31}$ values}
\label{phi31toV}
\citet{Jurcsik96} used $V$ band light curves to determine a relation between $\phi_{31}$ and metallicity. \citet{Dorfi99} and \citet{Deb10} established two independent linear relations of the form $\phi_{k1}^V = \alpha + \beta\phi_{k1}^I$ to transform the $I$ band Fourier parameters to $V$ band values. The values for $\alpha$ and $\beta$ shown in Table\,\ref{table_Fourier_parameter_LMC} are taken from Table\,3 of \citet{Deb10}. \\\noindent\hspace*{1em} 
\begin{table}
\caption{Parameters to transform the Fourier coefficient $\phi_{31}$ from the $I$ band to the $V$ band.}             
\label{table_Fourier_parameter_LMC}      
\centering                          
\begin{tabular}{c r r}        
\hline\hline                 
$\phi_{31}^V = \alpha + \beta\phi_{31}^I$ & $\alpha$ & $\beta$ \\    
\hline                        
\citet{Dorfi99} & $-0.039$ & 0.788 \\ 
\citet{Deb10} & 0.436 & 0.568 \\ 
\hline                                   
\end{tabular}
\end{table}
\citet{Deb10} used the observed light curves of the RR\,Lyrae stars of two globular clusters, with metallicities of $-2.16$\,dex \citep[M68,][]{Lee05} and $-1.39$\,dex \citep[M3,][]{Cohen05}, to calculate Fourier parameters. With these data they obtained values for $\alpha$ and $\beta$ that are significantly different from \citet{Dorfi99}, who used Fourier-decomposed theoretical light curve models of RR\,Lyrae stars to determine the relation. These models were matched with observational data from three globular clusters with metallicities of $-2.16 \pm 0.02$\,dex \citep[M68,][]{Lee05}, $-1.52 \pm 0.12$\,dex \citep[IC4499,][]{Hankey11} and $-1.18 \pm 0.02$\,dex \citep[NGC1851,][]{Carretta10}. \\\noindent\hspace*{1em} 
We transform the Fourier parameters provided by OGLE using the relations listed in Table\,\ref{table_Fourier_parameter_LMC}. The recalculated values of $\phi_{31}^V$ are inserted into the formula determined by \citet{Jurcsik96}: 
\begin{equation}
\label{Jurcsik96_FeH}
\mathrm{[Fe/H]} = -5.038 - 5.394P + 1.345\phi_{31}^V
\end{equation}
to redetermine the MDF of the MCs. All metallicities presented in this section are on the \citetalias{Jurcsik95} scale. Owing to the very small number of calibration clusters and the narrow metallicity range of \citet{Dorfi99} and \citet{Deb10}, we use this relation with a note of caution outside the metallicity regime defined by the globular clusters.
\subsection{LMC}
\label{phi31toV_LMC}
\begin{figure}
\centering 
 \includegraphics[height=0.30\textheight,width=0.47\textwidth]{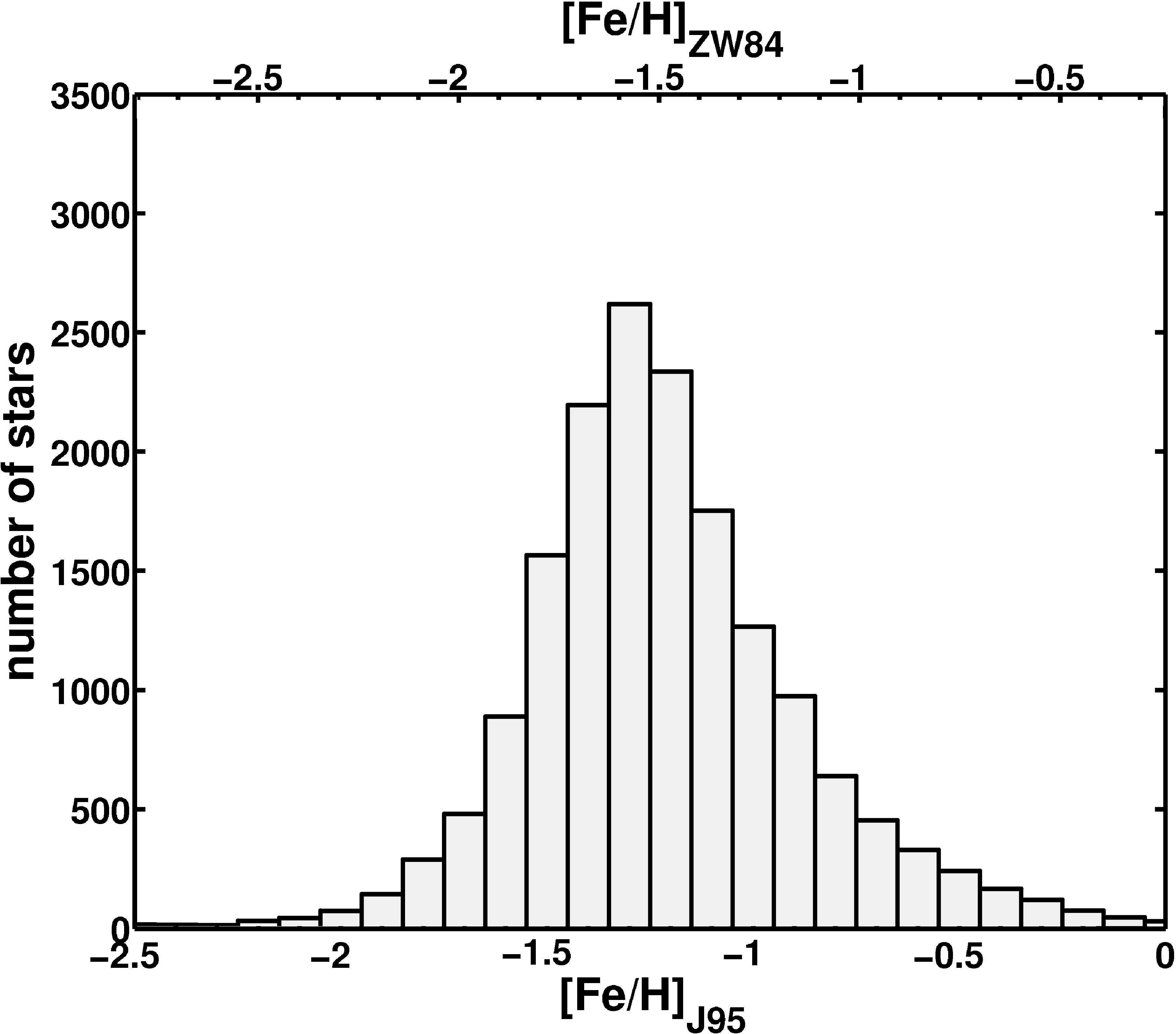}
\vspace{\floatsep}
 \includegraphics[height=0.30\textheight,width=0.47\textwidth]{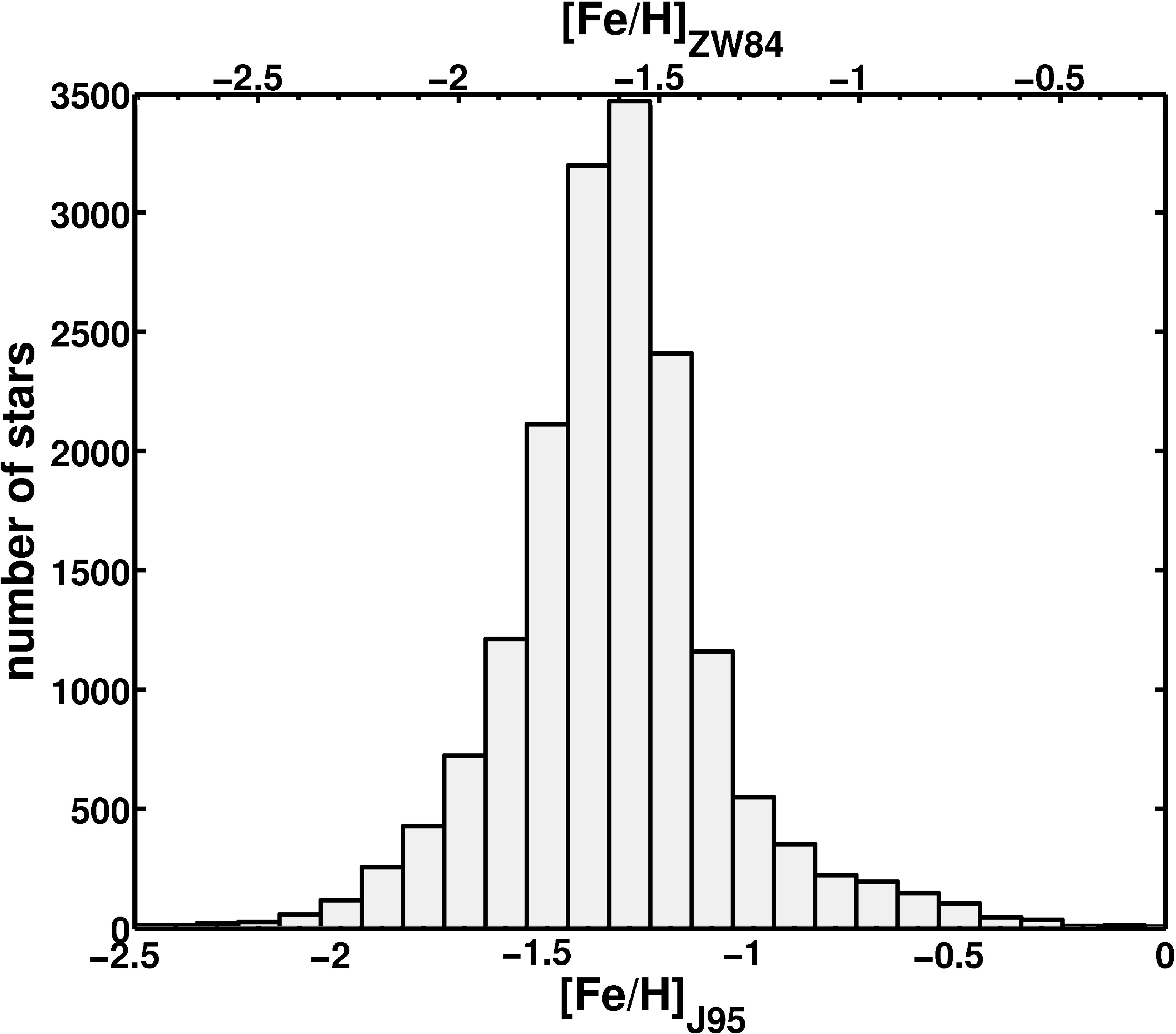}
 \caption{The upper panel shows the histogram of the LMC metallicity distribution obtained using the conversion of $\phi_{31}^{I}$ to $\phi_{31}^{V}$ by \citet{Dorfi99}, while the lower panel is obtained with the relation of \citet{Deb10}.} 
 \label{MDF_LMC_VI} 
\end{figure}

The relation by \citet{Dorfi99} leads to a very different metallicity distribution (upper panel in Figure\,\ref{MDF_LMC_VI}) than obtained in Figure\,\ref{MDF_LMC}. The resulting metallicity distribution is much broader with a standard deviation of 0.38\,dex, while the mean metallicity of $\mathrm{[Fe/H]}_{\mathrm{mean}/\mathrm{DF99}} = -1.20 \pm 0.38$\,dex is very similar to the value obtained with the \citet{Smolec05} relation. \\\noindent\hspace*{1em} 
Using coefficients from \citet{Deb10} to calculate metallicities of all RR\,Lyrae stars results in a metallicity distribution (lower panel in Figure\,\ref{MDF_LMC_VI}) similar in shape to that shown in Figure\,\ref{MDF_LMC}. The mean metallicity is comparable within the errors. With $\mathrm{[Fe/H]}_{\mathrm{mean}/\mathrm{DS10}} = -1.33 \pm 0.29$\,dex, we find a metallicity that is about 0.1\,dex lower than from the \citet{Smolec05} method. For this distribution a standard deviation of 0.29\,dex is obtained assuming a Gaussian distribution. \\\noindent\hspace*{1em} 
\subsection{SMC}
\label{phi31toV_SMC}
\begin{figure}
\centering 
 \includegraphics[height=0.30\textheight,width=0.47\textwidth]{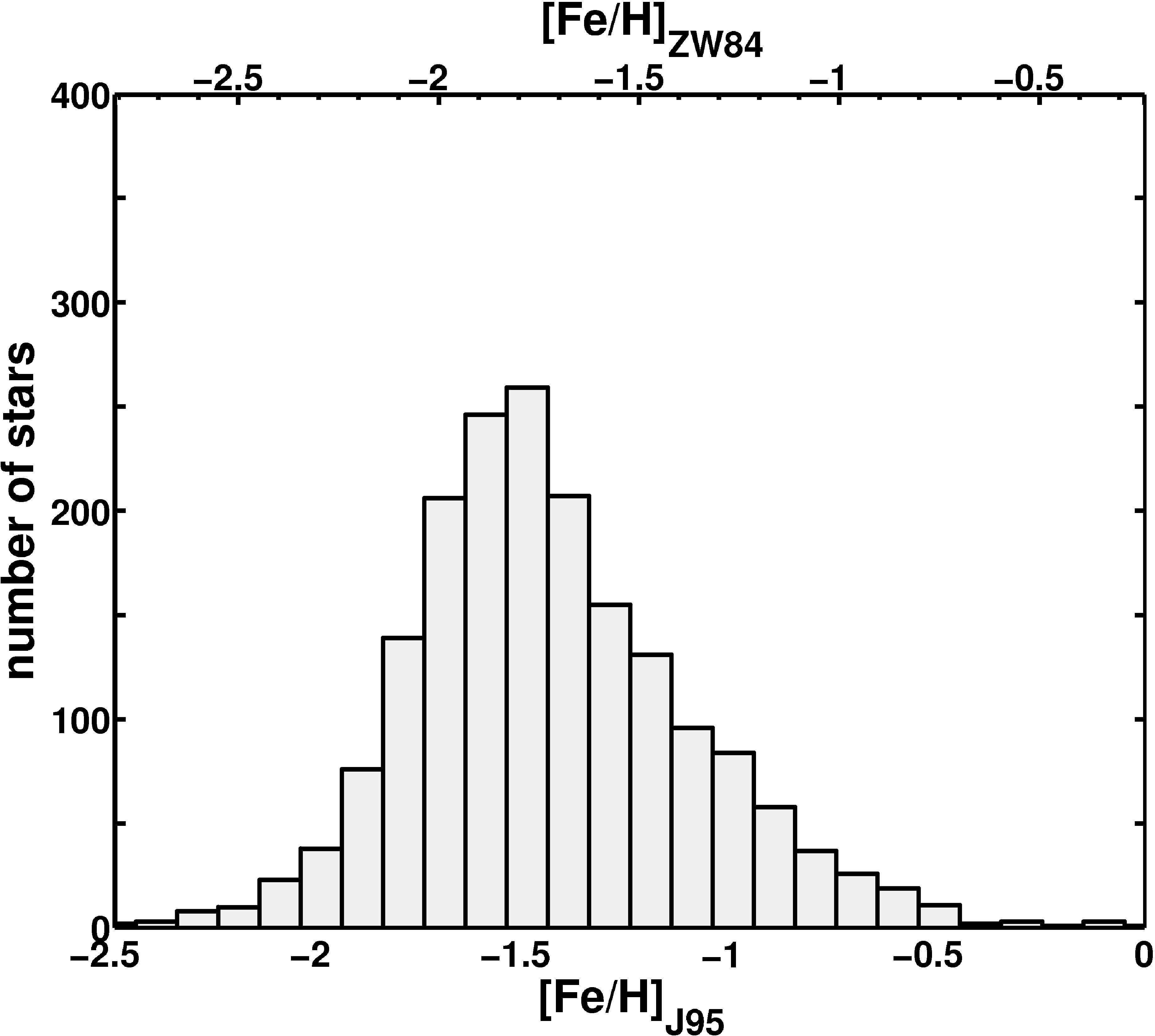}
\vspace{\floatsep}
 \includegraphics[height=0.30\textheight,width=0.47\textwidth]{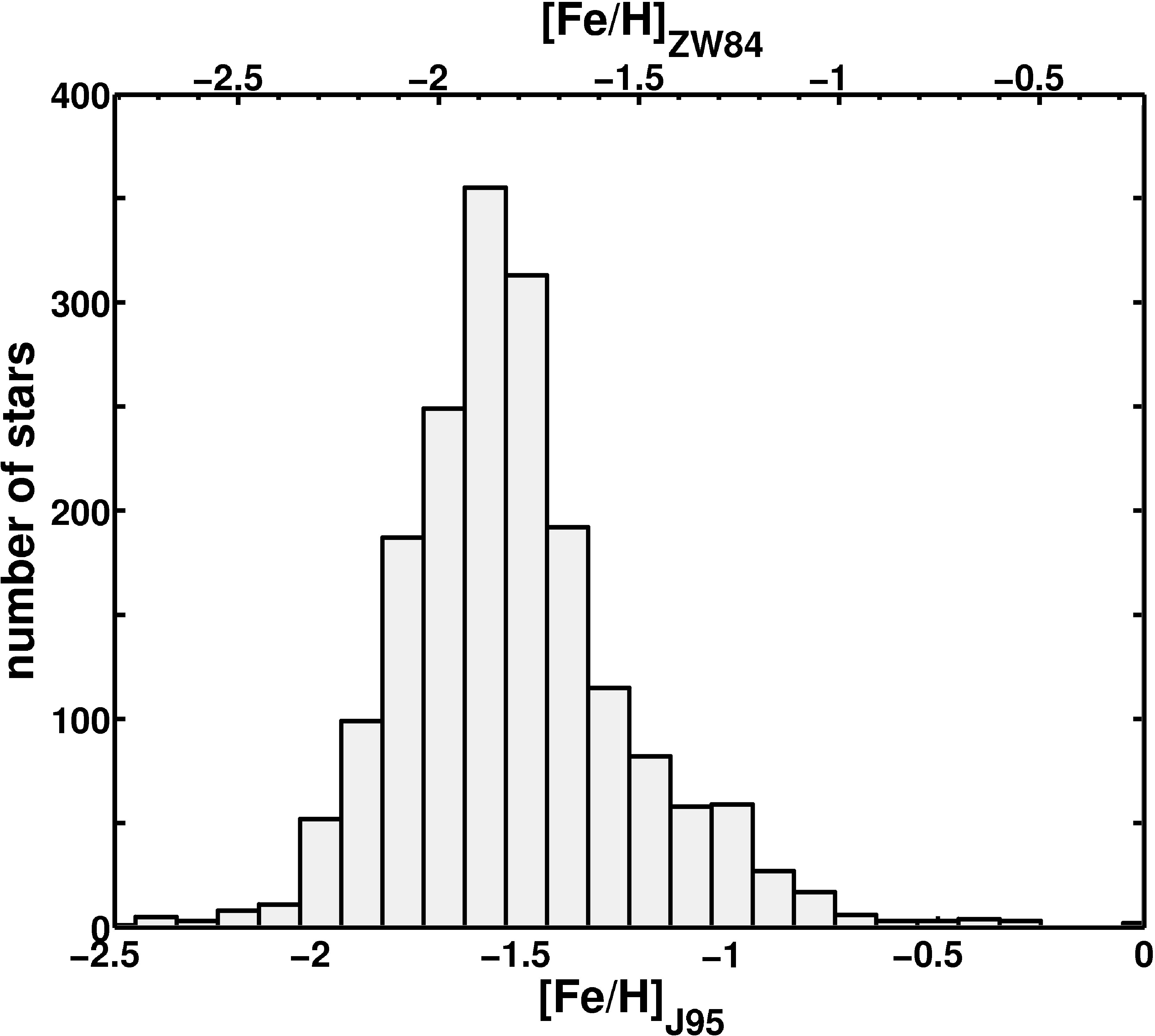}
 \caption{\citet{Dorfi99} conversion of the Fourier parameter $\phi_{31}^{I}$ to $\phi_{31}^{V}$ is used in the upper panel to determine the metallicity of the SMC RR\,Lyrae stars. In the lower panel the relation by \citet{Deb10} is applied to determine the metallicity distribution.} 
 \label{MDF_SMC_VI} 
\end{figure}

Using the \citet{Dorfi99} relation, the mean metallicity, on the \citetalias{Jurcsik95} scale, is $\mathrm{[Fe/H]}_{\mathrm{mean}/\mathrm{DF99}} = -1.43$\,dex with a standard deviation of 0.40\,dex. As with the LMC, the shape of the SMC metallicity distribution is much flatter and wider than with the relation by \citet{Smolec05}, while the mean metallicity is very similar (upper panel of Figure\,\ref{MDF_SMC_VI}).\\\noindent\hspace*{1em} 
The relation of \citet{Deb10} reveals a similar shape of the metallicity distribution compared with that based on the relation by \citet{Smolec05}. However, the metallicity is about 0.1\,dex lower with $\mathrm{[Fe/H]}_{\mathrm{mean}/\mathrm{DS10}} = -1.53$\,dex, while the standard deviation is 0.32\,dex. These results are shown in the lower panel of Figure\,\ref{MDF_SMC_VI} and are analogous to the outcome for the LMC. \\\noindent\hspace*{1em}

Overall, $72\%$ of the SMC metallicities determined with the relation by \citet{Deb10} and $21\%$ of the metallicities from the relation by \citet{Dorfi99} lie outside of the metallicity range defined by the calibrating clusters used to derive the relations. The relation by \citet{Smolec05} has not been calibrated for such low metallicities as the ones we find for some of our RR\,Lyrae stars, but includes a wider metallicity range. Especially in the metallicity interval of $-1.71\,\mathrm{dex} < \mathrm{[Fe/H]}_{\mathrm{J95}} < -1.30$\,dex, where most of the RR\,Lyrae stars are located in our sample, \citet{Smolec05} has a good coverage with 13 stars, thus justifying our use of \citet{Smolec05} relation as opposed to the other relations described here.

\section{Conclusions}
\label{Summary}

We derive photometric metallicities for 16949 RR\,Lyrae stars in the observed OGLE\,III field of the LMC using the Fourier decomposition of the light curves \citep{Jurcsik96}. On the \citetalias{Jurcsik95} metallicity scale we find a mean metallicity of $\mathrm{[Fe/H]}_{\mathrm{mean}} = -1.23$\,dex. This corresponds to a metallicity of $\mathrm{[Fe/H]}_{\mathrm{mean}} = -1.50$\,dex on the \citetalias{Zinn84} scale, which is in excellent agreement with the spectroscopic measurements of LMC RR\,Lyrae stars in the literature \citep{Borissova04, Borissova06, Gratton04}. Our results are also in excellent agreement with those of \citet{Smolec05}. The mean metallicity of his dataset, which is divided into Blazhko and non-Blazhko RR\,Lyrae stars, is $\mathrm{[Fe/H]}_{\mathrm{mean}} = -1.23$\,dex. The difference in the mean metallicities between the two types ($\Delta\mathrm{[Fe/H]} = 0.06$\,dex) is much smaller than the uncertainties associated with our method. We conclude that we would not detect a significant difference between the two types of RR\,Lyrae stars and therefore treat the stars independently of their type. Although star-to-star fluctuations in metallicity are observed on small scales, we find no evidence for a spatial metallicity gradient in the LMC. \\\noindent\hspace*{1em} 
For the SMC we find a mean metallicity of $\mathrm{[Fe/H]}_{\mathrm{mean}} = -1.42$\,dex on the \citetalias{Jurcsik95} scale based on 1831 RR\,Lyrae stars of type\,ab. This translates to a value of $\mathrm{[Fe/H]}_{\mathrm{mean}} = -1.70$\,dex on the scale of \citetalias{Zinn84}, and is in very good agreement with the spectroscopic determinations by \citet{Butler82} and the photometric metallicities by \citet{Kapakos11}. No metallicity gradient is observed in the old population of the SMC. \\\noindent\hspace*{1em} 
In the LMC, several old globular clusters with ages greater than 10\,Gyr are known. The metallicities of these clusters span at least one dex in [Fe/H] \citep[e.g,][]{Johnson06, Mucciarelli10}. This is smaller than the range of metallicities observed for Galactic halo globular clusters, but well within the range of what we found for the RR\,Lyrae field stars, although most of the old LMC globulars with spectroscopic metallicity determinations are more metal-poor than the peak of our MDF of the RR\,Lyrae stars. In the SMC only one globular cluster, NGC\,121, is known \citep{Glatt08a}, and has a metallicity of $-1.46 \pm 0.1$\,dex on the \citetalias{Zinn84} scale \citep{DaCosta98}, well within the range of metallicities of SMC field RR\,Lyrae stars. The peak of our SMC RR\,Lyrae MDF is slightly more metal-poor than the metallicity of NGC\,121. The SMC in fact shows a wide range of metallicities at any given age \citep[e.g.,][]{Kayser07, Glatt08b}.  \\\noindent\hspace*{1em} 
Large metallicity spreads have also been found in most Local Group dwarf spheroidal (dSphs), usually exceeding one dex even when focusing on their old (red giant branch) populations \citep[e.g.,][]{Grebel03}. For the old population of the LMC we find an intrinsic metallicity spread of $\sigma = 0.24$\,dex (in terms of one standard deviation of the distribution), and $\sigma = 0.27$\,dex for the SMC. The full range of metallicities of the RR\,Lyrae stars exceeds 1\,dex in [Fe/H] and is difficult to constrain owing to the limits of the validity range of our calibration.  As pointed out already, in neither of the two Clouds do we find a statistically significant trend in metallicity with position despite the spread. \\\noindent\hspace*{1em} 
While we cannot assign specific ages to our individual RR\,Lyrae stars, we suggest that this spread was caused by inhomogeneous, localized enrichment, which may happen on relatively short time scales as discussed by \citet{Marcolini08}. Detailed spectroscopic element abundance ratios measured for red giants in old LMC globular clusters indicate star formation from gas enriched by supernovae of Type\,II with prominent contributions from the r-process \citep{Mucciarelli10}.  The $\alpha$ enhancement is comparable to Galactic halo stars (and thus inconsistent with the [$\alpha$/Fe] ratios found in the old populations of many of the Galactic dSph and ultra-faint dSph stars; e.g., \citet{Koch08} and \citet{Aden11}. \citet{Johnson06} find differences between LMC globular cluster element abundance ratios both in comparison to dSphs and to Galactic globular clusters, emphasizing that the LMC experienced a different star formation history than the Galactic old populations.  No such detailed, age-dateable information is available for the SMC thus far. \\\noindent\hspace*{1em} 
If we qualitatively compare our RR\,Lyrae MDFs to MDFs derived spectroscopically for Galactic dSphs (see, e.g., Figure\,13 in \citealt{Koch06}; Figure\,7 in \citealt{Koch07a}; Figure\,12 in \citealt{Koch07b}, and Figure\,1 in \citealt{Kirby11}) we find that the MDFs of the Clouds do not show the steep drop-off at the metal-rich end found in some dSphs.  In the chemical evolution models for dSphs of \citet{Lanfranchi04}, such a steep decline is caused by gas outflows, which then do not seem to have played a defining role in the evolution of the old population of the Magellanic Clouds. This may not be surprising given the presumably much deeper potential wells of the Clouds.  The shape of the MDFs of the MCs' old populations nevertheless resembles the shape of the MDFs of several Galactic dSphs that, according to \citet{Kirby11}, agree well with the gas-infall models of \citet{Lynden-Bell75} and \citet{Pagel97}, suggesting gas accretion from the intergalactic medium at early times.  Note that since many of the Galactic dSphs contain predominantly old populations, we are comparing stars formed during similar epochs here.  \\\noindent\hspace*{1em} 
Although the MDFs of the RR\,Lyrae stars in the MCs reveal a considerable metallicity spread, we do not see evidence for a large population of stars with [Fe/H] $< -2$\,dex.  We caution that the calibration of the Fourier decomposition method does not extend to low metallicities and that it has been suggested that this method overestimates metallicities at low [Fe/H] values (\citealt{Dekany09}, but see also \citealt{Arellano11}). Additionally the number of RR\,Lyrae stars with low metallicities could be influenced by the shift of the instability strip to lower temperatures with lower metallicities. This could result in fewer horizontal branch stars experiencing the RR\,Lyrae phase \citep{Yoon02}. However, very metal poor globular clusters (e.g. NGC 2419 and NGC 7078) also contain a considerable fraction of RR\,Lyrae stars \citep{Catelan09}. On the other hand, there are a number of metal-poor globular clusters both in the Milky Way and in the LMC that are dominated by blue horizontal branch morphologies and that essentially do not contain any RR\,Lyrae stars \citep{Catelan09}. Hence it is conceivable that the apparent lack of metal-poor RR\,Lyrae stars is also caused by this evolutionary effect. Nonetheless, the apparent absence of a substantial, very metal-poor old population suggests that the early enrichment in the MCs proceeded rapidly and efficiently. The MCs show a similar lack of metal-deficient stars as the Galactic dSphs and the Galactic halo (also known as the ``G dwarf problem''), suggesting that the majority of the old stars formed from pre-enriched gas. Spectroscopic studies of our RR\,Lyrae metal-poor candidates would be desirable to confirm their metallicities.   \\\noindent\hspace*{1em} 
The differences in the mean metallicities of the old populations in the LMC and SMC makes a common origin of these two galaxies from a single protogalactic fragment unlikely, unless such a putative progenitor would have been separated into two objects very early on. Recently, \citet{Yang10} proposed that the LMC might have been ejected from M31 in a major merger event. The comparatively high metallicity of old globular clusters in M31 \citep{Caldwell11} as compared to the properties of the MDF of the old LMC population seems to argue against this scenario. 

%

\begin{acknowledgements}
We thank our referee Prof. Horace Smith for very useful comments. We are grateful to the OGLE collaboration for making their data publicly available. R.\,Haschke is obliged to Katharina Glatt and Katrin Jordi for proof reading and useful discussions that helped to improve the paper. This work was partially supported by Sonderforschungsbereich SFB\,881 ``The Milky Way System'' (subproject A2) of the German Research Foundation (DFG).  \end{acknowledgements}

\bibliography{Bibliography}
\bibliographystyle{apj}

\end{document}